\documentclass[aps,prl,reprint,twocolumn,superscriptaddress]{revtex4-2}

\usepackage[T1]{fontenc}
\usepackage{graphicx}
\usepackage[hidelinks,colorlinks=true,linkcolor=blue,citecolor=blue,urlcolor=blue]{hyperref}
\usepackage{amsmath}
\usepackage{bbold}
\usepackage{placeins}
\usepackage{multirow}
\usepackage[table,xcdraw]{xcolor}
\newcommand{\ket}[1]{| #1 \rangle}
\newcommand{\acomment}[1]{}

\begin{document}
\raggedbottom


\title{Demonstration of an always-on exchange-only spin qubit}

\author{Joseph D. Broz}
\email[]{jdbroz@hrl.com}
\affiliation{HRL Laboratories, LLC, 3011 Malibu Canyon Road, Malibu, California 90265, USA}
\author{Jesse C. Hoke}
\affiliation{HRL Laboratories, LLC, 3011 Malibu Canyon Road, Malibu, California 90265, USA}
\author{Edwin Acuna}
\affiliation{HRL Laboratories, LLC, 3011 Malibu Canyon Road, Malibu, California 90265, USA}
\author{Jason R. Petta}
\email[]{jpetta1@hrl.com}
\affiliation{HRL Laboratories, LLC, 3011 Malibu Canyon Road, Malibu, California 90265, USA}
\affiliation{Department of Physics and Astronomy, University of California -- Los Angeles, Los Angeles, California 90095, USA}
\affiliation{Center for Quantum Science and Engineering, University of California -- Los Angeles, Los Angeles, California 90095, USA}

\date{\today}

\begin{abstract}
In conventional exchange-only (EO) spin qubit demonstrations, quantum gates have been implemented using sequences of individually pulsed pairwise exchange interactions with only one exchange coupling active at a time. Alternatively, multiple non-commuting exchange interactions can be pulsed simultaneously, reducing circuit depths and providing protection against leakage. We demonstrate high-fidelity quantum control of an always-on exchange-only (AEON) qubit, operated using simultaneous exchange pulses in a triangular quantum dot (QD) array. We use blind randomized benchmarking to characterize the performance of the full AEON single-qubit Clifford gate set, achieving an average Clifford gate fidelity $F_{\rm C1}$ = 99.86\%. Extensions of this work may enable more efficient EO two-qubit entangling gates as well as the implementation of native $i$-Toffoli gates in Loss-DiVincenzo single-spin qubits.
\end{abstract}

\maketitle


\section{Introduction}

Gate-defined semiconductor QDs are an attractive platform for scalable quantum computing owing to their small size, compatibility with existing semiconductor fabrication techniques, and potential for operation at temperatures above 1 K \cite{burkard2023,Ansaloni2020,ha2022,Zwerver2022,Neyens2024,Petit2020,Yang2020}. In QDs, quantum information is typically stored in the spins of confined electrons or holes, which benefit from long coherence times \cite{Veldhorst2014,Muhonen2014} and the availability of a fast, electrically controlled exchange interaction for coupling nearest neighbor spins \cite{loss1998, petta2005}. However, single-spin qubit control is difficult, requiring either precise engineering of local magnetic fields or the use of materials with strong spin-orbit coupling, both of which add to design complexity \cite{tokura2006,Pioro2008,Watzinger2018}.

The EO qubit was developed to circumvent the need for single-spin control by encoding a single qubit in a subspace of three physical spins, which support universal quantum control through the exchange interaction alone \cite{DiVincenzo2000}. The encoded gates are typically implemented using sequences of pulsed pairwise exchange interactions, where no more than two spins interact at any given time \cite{laird2010, Medford2013, Andrews2019, Weinstein2023}. This serial mode of operation is chosen for its practical simplicity, as qubit control is characterized by a single parameter --- a time-integrated exchange energy $J_{i,j}(t)$ coupling spins $i$ and $j$ --- which is easy to tune and calibrate \cite{Andrews2019}. However, this simplicity comes at the cost of large gate depths, with recently demonstrated single-qubit gates requiring up to four exchange pulses and two-qubit gates as many as twenty-eight \cite{Andrews2019, Weinstein2023}.  EO gates can be designed to further mitigate leakage error sources, but at the cost of even longer sequence lengths, such as the 16-pulse single-qubit gates designed to dynamically decouple magnetic and charge noise \cite{hickman2013}, or the 44-pulse leakage-controlled two-qubit gates which suppress spreading of leakage errors \cite{Weinstein2023}.

The AEON qubit is a variant of the EO qubit that uses simultaneously pulsed pairs of exchange couplings to construct quantum gates \cite{Shim2016}. A key advantage of the AEON qubit is significantly shorter gate depths, with single-qubit gates requiring no more than two simultaneous exchange pulses \cite{Shim2013SinglequbitGI} and two-qubit gates as little as one \cite{Weinstein2005,doherty2013_2qubit_rx_gates}. Moreover, these gates are naturally protected from leakage due to an induced energy gap between the qubit and leakage subspaces \cite{hung2014,Weinstein2005,doherty2013_2qubit_rx_gates}.

Here we report full control of an AEON qubit in a Si/SiGe triangular QD array \cite{acuna2024}. Sensitivity to charge noise is mitigated by operating exchange at a ``double sweet spot'' (DSS), which is first-order insensitive to fluctuations of the chemical potentials of all three confined electrons \cite{Russ2016,Shim2016}. We implement a calibration procedure for AEON qubit gates that simultaneously calibrates both the axis and angle of rotation. With this procedure, we realize a complete set of gates and extract a single-qubit Clifford gate fidelity $F_{C1}$ = 99.86\% via blind randomized benchmarking (BRB). Our results show that AEON qubit performance is on par with state-of-the-art EO qubits operated using sequential exchange pulses \cite{Andrews2019}.


\section{Device Physics}
\label{sec:device}
\begin{figure*}[th!]
	\includegraphics[width=0.99\textwidth]{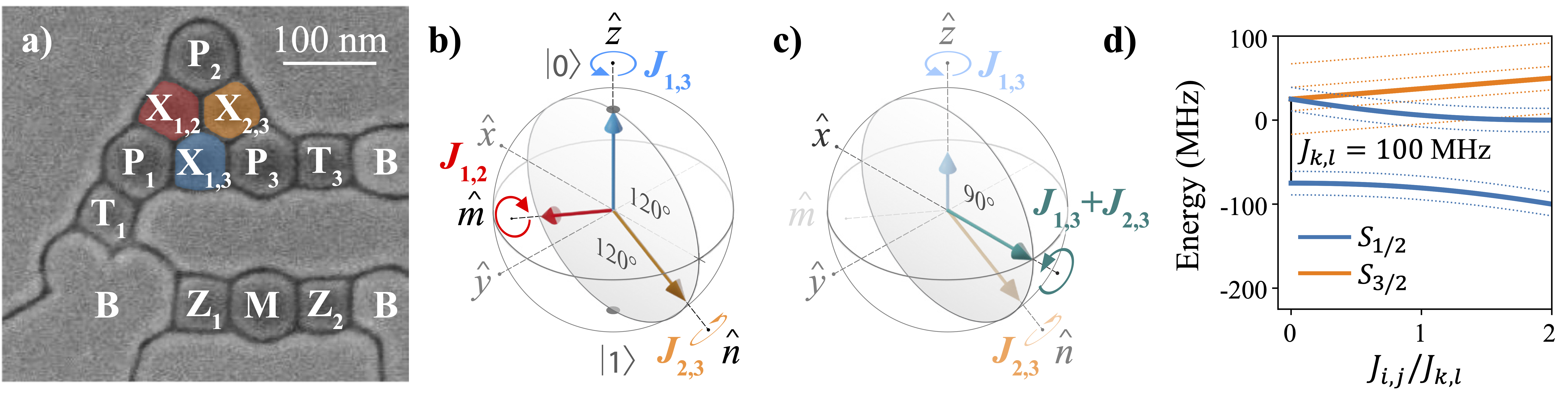}
	\caption{\label{fig:figure1} \textbf{Device operation}. (a) Scanning electron microscope image of a triangular QD similar to the one measured. (b) During conventional EO qubit operation, a single exchange coupling generates a rotation about one of three non-orthogonal axes ($\hat{z}$, $\hat{n}$, or $\hat{m}$). (c) More efficient control sequences can be implemented if two exchange couplings are simultaneously activated. For example, an $x$-axis rotation can be implemented by applying simultaneous voltage pulses to the $\textrm{X}_{1,3}$ and $\textrm{X}_{2,3}$ gates. (d) Eigenergies of Eq.~\ref{eq:H_exc spin operators} plotted as a function of $J_{i,j}$ with $J_{k,l}$~=~$2\pi$~$\times$~100~MHz ($i,j\ne k,l$), $J_{i,l}$~=~0, and $B = 0$. Solid blue curves correspond to the $S\mathbin{=}1/2$ doublets and the solid orange curve to the $S\mathbin{=}3/2$ leakage space quadruplets. When both $J_{i,j}$ and $J_{k,l}$ are nonzero, the $S\mathbin{=}1/2$ and $S\mathbin{=}3/2$ subspaces are energetically isolated, suppressing leakage out of the qubit subspace. A nonzero global magnetic field (dotted energy levels), lifts most of the remaining degeneracies. To preserve leakage protection in the presence of a global magnetic field, simultaneous exchange energies should be chosen to avoid level crossings.}
\end{figure*}

Measurements are performed on a triangular triple QD device consisting of a single layer of etch-defined gate electrodes fabricated on top of an isotopically enriched $^{28}$Si/SiGe heterostructure (800 ppm $^{29}$Si) \cite{ha2022, acuna2024}. A scanning electron microscope image of a nominally identical device is shown in Fig.~\hyperref[fig:figure1]{1(a)}. Voltages applied to plunger gates ($\textrm{P}_i$) form QDs in the underlying $^{28}$Si quantum well, each tuned to confine a single electron. Voltages applied to exchange gates ($\textrm{X}_{i,j}$) control the exchange coupling between neighboring QDs. In practice, devices are tuned using ``virtual gates,'' linear combinations of physical gate voltages that selectively control key dot parameters while compensating for device cross-capacitances \cite{Hensgens2017fermi-hubbard,mills2019,hsiao2020}. Specifically, the virtual plunger gate voltage $\tilde{V}_{\textrm{P}_i}$ controls the chemical potential $\epsilon_i$ of dot $i$ and the virtual exchange gate voltage $\tilde{V}_{\textrm{X}_{i,j}}$ controls the tunnel coupling $t_{i,j}$ between dots $i$ and $j$ (see Methods for more details). The remaining gate electrodes are used to load electrons into the dots from charge reservoirs (\textrm{B}, \textrm{T$_i$}) and to form a dot charge sensor (\textrm{M}, $\textrm{Z}_i$) \cite{blumoff2022}. 

\begin{figure*}
\centering
\includegraphics[width=0.99\textwidth]{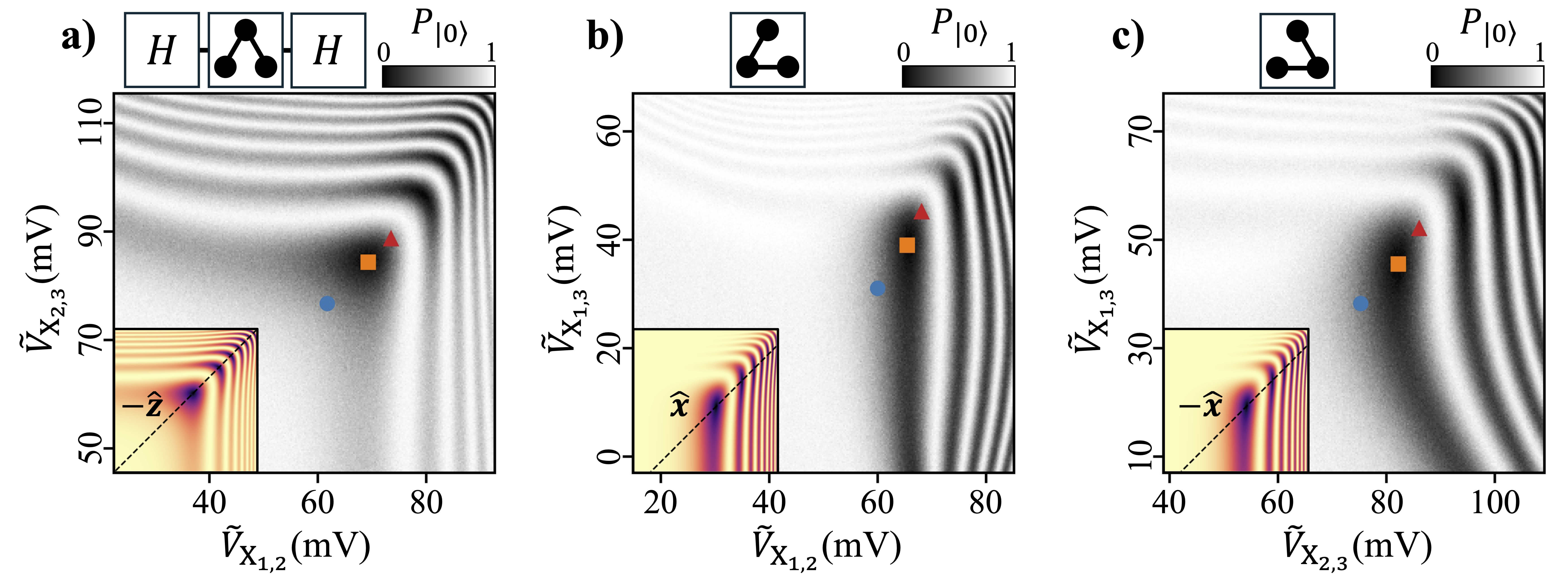}
\caption{\label{fig:figure2} \textbf{Quantum control with simultaneous exchange couplings}. (a)--(c) The probability $P_{\ket{0}}$ of measuring the encoded $\ket{0}$ state following a 10 ns 2-$J$ exchange pulse as a function of pulse amplitude for all pairwise combinations of virtual exchange gates. The colored markers identify the locations of $\pi/2$ (blue circles), $\pi$ (orange squares), and $3\pi/2$ (red triangles) rotations. The target axes about which these rotations are performed are the same as those labeled in the insets. For the data in (a), 1-$J$ Hadamard ($H$) rotations are applied before and after the 2-$J$ exchange pulse. Insets: Simulations of the ideal two-level system evolution as given by Eq.~\ref{eq:qubit rotation} and assuming an exponential dependence of the exchange energy on virtual exchange gate voltages. 
}
\end{figure*}

The Hamiltonian for the three singly-occupied exchange-coupled QDs is given by:
\begin{equation} \label{eq:H_exc spin operators}
    \hat{H}(t) = J_{1,2}(t)\hat{\mathbf{S}}_1\cdot\hat{\mathbf{S}}_2 + J_{2,3}(t)\hat{\mathbf{S}}_2\cdot\hat{\mathbf{S}}_3 + J_{1,3}(t)\hat{\mathbf{S}}_1\cdot\hat{\mathbf{S}}_3,
\end{equation}
\noindent where $\hat{\mathbf{S}}_i$ are the dot spin operators and we set $\hbar$ = 1 such that $J_{i,j}$ can be interpreted as either an energy or angular frequency. We introduce the notation $k$-$J$ to describe the number $k$ of nonzero $J_{i,j}$ contained in $\hat{H}$. EO or AEON qubit control is then distinguished by the restriction of $\hat{H}$ to 1-$J$ or 2-$J$ exchange, respectively.

The qubit is encoded in the collective three-electron spin state of the array, which occupies an eight-dimensional Hilbert space \cite{DiVincenzo2000}. In terms of the total (three-electron) spin $S$ and its projection $m_S$, this Hilbert space decomposes into a $S\mathbin{=}3/2$ quadruplet ($m_S\mathbin{=}\pm 1/2,\pm 3/2$) and two $S\mathbin{=}1/2$ doublets ($m_S\mathbin{=}\pm 1/2$). The exchange interaction, which conserves both $S$ and $m_S$, only couples states within the two doublets. The qubit is encoded by defining $\ket{0} = \ket{S_{13}\mathbin{=}0,S\mathbin{=}1/2,m_S}$ and $\ket{1} = \ket{S_{13}\mathbin{=}1,S\mathbin{=}1/2,m_S}$, where $S_{13}$ is the combined spin of dots 1 and 3 and $m_S\mathbin{=\pm 1/2}$ acts as an extra ``gauge'' degree of freedom \cite{burkard2023}. The state $\ket{0}$ corresponds to a singlet between dots 1 and 3, with the uncoupled spin of dot 2 determining the qubit's gauge. We use standard Pauli spin blockade techniques \cite{petta2005} to initialize the qubit into the singlet state $\ket{0}$ and to map the occupation of $\ket{0}$ onto the charge configuration, which can then be measured using the dot charge sensor \cite{blumoff2022}.  The specific value of $m_S$ is randomly assigned during initialization and is left unresolved by measurement.

Expressed in the qubit basis, Eq.~\ref{eq:H_exc spin operators} takes the form: 
\begin{equation} \label{eq;H_exc qubit operators}
    \hat{H}(t) = -\frac{1}{2}[\sqrt{3}J_-(t)\hat{\sigma}_x + (J_{1,3}(t) - J_+(t))\hat{\sigma}_z],
\end{equation}
\noindent where $\hat{\sigma}_i$ are the standard Pauli operators, and we define $J_+=(J_{1,2} + J_{2,3})/2$ and $J_-=(J_{1,2} - J_{2,3})/2$. In the Bloch sphere picture, time-evolution under $\hat{H}$ corresponds to a rotation with angle $\theta$ about an axis $\hat{r}\mathbin{=}(\textrm{cos}(\varphi), 0, \textrm{sin}(\varphi))$ in the $xz$-plane, as described by the unitary operator
\begin{align} 
    \hat{R}_{\varphi}(\theta) &= \mathcal{T}\textrm{exp}[-i\int_0^{\tau} \hat{H}(t)dt]\\
    &\approx \textrm{cos}(\theta/2) - i\textrm{sin}(\theta/2)[r_x\hat{\sigma}_x + r_z\hat{\sigma}_z],
    \label{eq:qubit rotation}
\end{align}
where $\mathcal{T}$ is the time-ordering operator, $\theta=\int_0^\tau \Omega(t) dt$, $\Omega = \sqrt{3J_-^2 + (J_{1,3} - J_+)^2}$, $r_x = \cos(\varphi) = \sqrt{3}J_-/\Omega$, and $r_z = \sin(\varphi) = (J_{1,3} - J_+) / \Omega$.
The approximation in Eq.~\ref{eq:qubit rotation} is exact only when $[\hat{H}(t),\hat{H}(t')]=0$ for all times $t$ and $t'$, which requires that the ratios of nonzero $J_{i,j}$ remain constant over the duration of the exchange pulses. This condition is trivially met for 1-$J$ exchange, which drives rotations about one of three non-orthogonal axes separated by $120^\circ$ in the $xz$-plane [see Fig.~\hyperref[fig:figure1]{1(b)}].  We label these as $\hat{z}=(0,0,1)$, $\hat{n}=-(\sqrt{3},0,1)/2$, and $\hat{m}=(\sqrt{3},0,-1)/2$. By interleaving rotations about any two of these axes, an arbitrary single-qubit gate can be constructed using at most four 1-$J$ exchange pulses \cite{Andrews2019}. 

In contrast, 2-$J$ exchange allows for rotations about an arbitrary axis in the $xz$-plane [see Fig.~\hyperref[fig:figure1]{1(c)}]. Using 2-$J$ exchange, any single-qubit gate can be implemented using no more than two pulses \cite{Shim2013SinglequbitGI}. Because the two constituent interactions are non-commuting, 2-$J$ qubit operation generally requires full temporal control of the voltage waveforms. However, we verify through experiment that even without implementing this level of control Eq.~\ref{eq:qubit rotation} still provides an accurate description of 2-$J$ operation. 

Due to its encoding in a Zeeman doublet, the qubit's dynamics are invariant to global magnetic fields \cite{Lidar1998,DiVincenzo2000,kempe2001,Fong2011}. However, fluctuating magnetic field gradients generated by nearby spinful nuclei can induce decoherence and leakage out of the qubit subspace \cite{ladd2012,Kerckhoff2021}. We mitigate leakage by applying a small global magnetic field $B \approx$~3 mT to suppress transitions that do not conserve $m_S$, leaving a single (gauge-dependent) leakage state, $\ket{1, 3/2, m_S}$, spin-degenerate with the qubit subspace. 1-$J$ exchange partially breaks this remaining degeneracy by introducing an energy gap between states with $S_{13}$~=~0 and $S_{13}$~=~1. On the other hand, 2-$J$ exchange energetically separates the $S\mathbin{=}1/2$ qubit and the $S\mathbin{=}3/2$ subspaces, highly suppressing leakage [see Fig.~\hyperref[fig:figure1]{1(d)}].


\section{Quantum Control with Simultaneous Exchange Couplings}

\begin{figure*}
\centering
\includegraphics[width=0.99\textwidth]{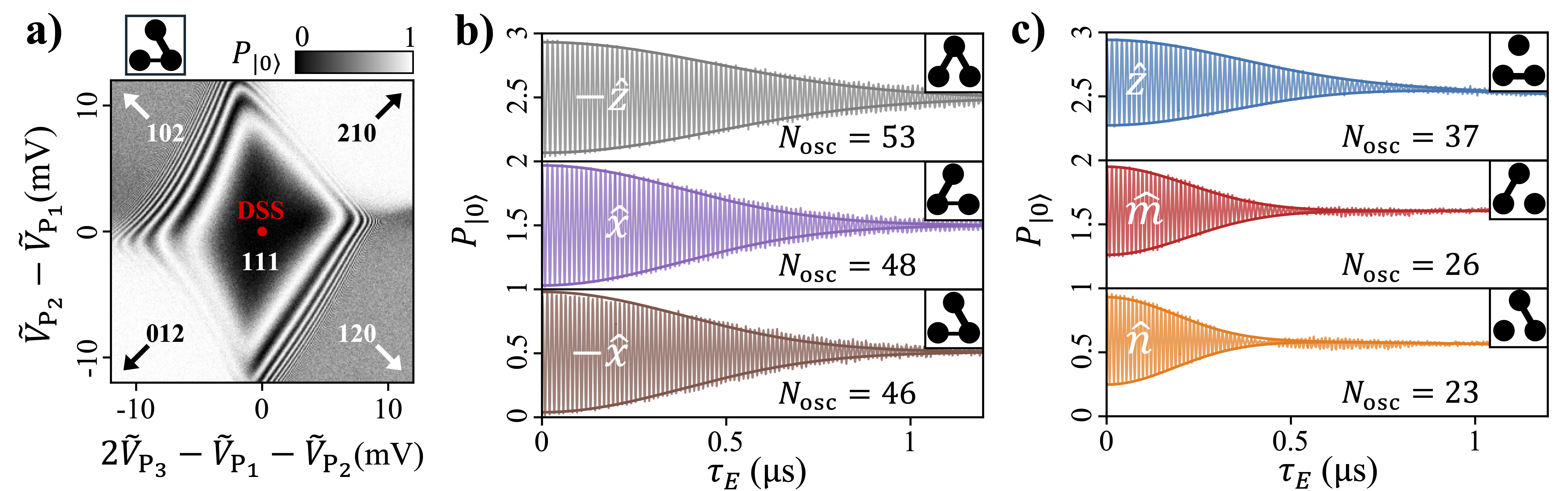}
\caption{\label{fig:figure3} \textbf{Coherent control of the AEON qubit}. (a) The 2-$J$ DSS, where the dependence of the exchange energy on dot chemical potentials vanishes to first order. By performing plunger gate sweeps, we identify the DSS as the point in bias space where $P_{\ket{0}}$ exhibits minimal sensitivity to \textrm{P} gate voltages (red circle). The linear combination of plunger gate voltages swept along each axis are chosen to be mutually orthogonal with each other and the common mode voltage $\tilde{V}_{P_1} + \tilde{V}_{P_2} + \tilde{V}_{P_3}$. Tuples denote the locations of (or directions towards) the nearest relevant charge configurations (e.g.~111 denotes the equilibrium state where a single electron is confined in each dot). (b)--(c) Measured 1- and 2-$J$ exchange oscillations with the oscillation frequency, $\Omega$, tuned to approximately $2 \pi$~$\times$~$80$ MHz.  The white labels denote the targeted axis of rotation and successive curves are offset by 1 on the $y$-axis for clarity. $N_{\textrm{osc}}$ is extracted by fitting each data set to a Gaussian decay envelope (darker lines). We note that pre/post $\pi$-pulses are applied about the $\hat{n}$-axis when performing 1-$J$ $\hat{z}$-rotations (blue curve) and pre/post Hadamard rotations are applied when performing 2-$J$ -$\hat{z}$-rotations (gray curve). Insets: Weighted solid lines indicate relative exchange gate pulse amplitudes.
}
\end{figure*}

To characterize the 2-$J$ exchange landscape, we sweep the amplitudes of two virtual exchange gate voltages applied simultaneously during a 10 ns pulse, producing the ``fingerpinch'' plots shown in Fig.~\ref{fig:figure2} \cite{heinz2024}. The data reveal oscillations of the probability $P_{\ket{0}}$ of measuring the encoded $\ket{0}$ state as a function of exchange gate voltages, as expected from Eq.~\ref{eq:qubit rotation}. The insets in Fig.~\ref{fig:figure2} show simulations of the corresponding ideal two-level system evolution, assuming an uncoupled exponential dependence of exchange energy $J_{i,j}$ on the virtual exchange gate voltage $\tilde{V}_{\textrm{X}_{i,j}}$. While the measured data qualitatively resembles the simulations, nonlinear and, in some cases, non-monotonic deviations are evident due to a complicated cross-dependence of the 2-$J$ exchange interaction on gate voltages \cite{coherent_multispin_exchange}. 

In principle, the strength of a 2-$J$ exchange interaction depends on both the interdot tunnel couplings and the dot chemical potentials. In a triple-dot system, the chemical potentials are typically parameterized using a common mode and two differential modes \cite{Friesen2015}. Exchange is insensitive to the common mode, defined as $\bar{\epsilon}=\epsilon_1+\epsilon_2+\epsilon_3$, but sensitive to the differential modes: the tilt detuning, $\epsilon_t=(\epsilon_2-\epsilon_1)/2$, and the dimple detuning, $\epsilon_d=\epsilon_3 - (\epsilon_1+\epsilon_2)/2$. However, it is possible to operate 2-$J$ exchange at a DSS where the sensitivity to the two differential modes vanishes to first order, as demonstrated in Fig.~\hyperref[fig:figure3]{3(a)}, thereby reducing decoherence due to charge noise \cite{Friesen2015,Shim2016,Russ2016,Malinowski2017}. 

Operating at the DSS, we characterize coherence by measuring the qubit's time-evolution during 2-$J$ exchange pulses, as plotted in Fig.~\hyperref[fig:figure3]{3(b)}. We quantify coherence by the number of exchange oscillations $N_{\textrm{osc}}$ that occur before the amplitude decays to 1/$e$ of its initial value. For comparison, we also plot the measured time-evolution for 1-$J$ exchange in Fig.~\hyperref[fig:figure3]{3(c)} (see Supplementary Information for 1-$J$ fingerprint plots and associated operating points). Unexpectedly, the $N_{\textrm{osc}}$ for 2-$J$ exchange  is significantly greater than for 1-$J$ exchange. The improvement is not a direct consequence of operation at the DSS, which should result in comparable $N_{\textrm{osc}}$ for 1- and 2-$J$ exchange. More detailed device modeling may be helpful for understanding the coherence enhancement during 2-$J$ operation.

\section{Gate Calibration and Benchmarking}

To calibrate a 2-$J$ exchange rotation $\hat{R}_{\varphi}(\theta)$, both the rotation axis $\varphi$ and angle $\theta$ must be tuned.  We directly optimize the fidelity of a composite pulse sequence designed to be simultaneously sensitive to both 2-$J$ rotation parameters. The composite sequence is defined as $\hat{U}(N) = \hat{U}_{\textrm{ax}}^N\hat{U}_{\textrm{ang}}^N$ and reduces to the identity $\hat{\mathbb{1}}$ when the 2-$J$ rotation is perfectly calibrated. This occurs when $\hat{R}_{\varphi}(\theta)=\hat{R}_{\varphi^*}(\theta^*)$ for the target rotation angles $\varphi^*$ and $\theta^*$. The two components of $\hat{U}$ differ in their sensitivity: $\hat{U}_{\textrm{ax}}$ is primarily sensitive to errors in $\varphi$ and $\hat{U}_{\textrm{ang}}$ to errors in $\theta$. Explicitly, we define $\hat{U}_{\textrm{ang}} = \hat{R}_{\varphi}^{2q}(\theta)$ and $\hat{U}_{\textrm{ax}} = [\hat{R}_{\eta}(\chi)\hat{R}_{\varphi}^q(\theta)]^{2}$, where $q$ is a positive integer chosen such that $q\theta =s\pi$ for some odd integer $s$, and $\hat{R}_{\eta}(\chi)$ is a pre-calibrated rotation with $\chi = \pi$ and $\eta = \varphi^* \pm \pi/2$. Similar sequences are used in the context of robust phase estimation and gate set tomography to amplify deviations from target gate parameters \cite{kimmel2015rpe, Nielsen2021gatesettomography}. Finally, to map the fidelity $F(\hat{U}, \hat{\mathbb{1}})$ of $\hat{U}$ relative to the identity onto the measured probability $P_{\ket{0}}$, we twirl $\hat{U}$ over the set of 1-$J$ single-qubit Clifford gates \cite{Magesan2012}. 

\begin{figure}
\centering
\includegraphics[width=0.49\textwidth]{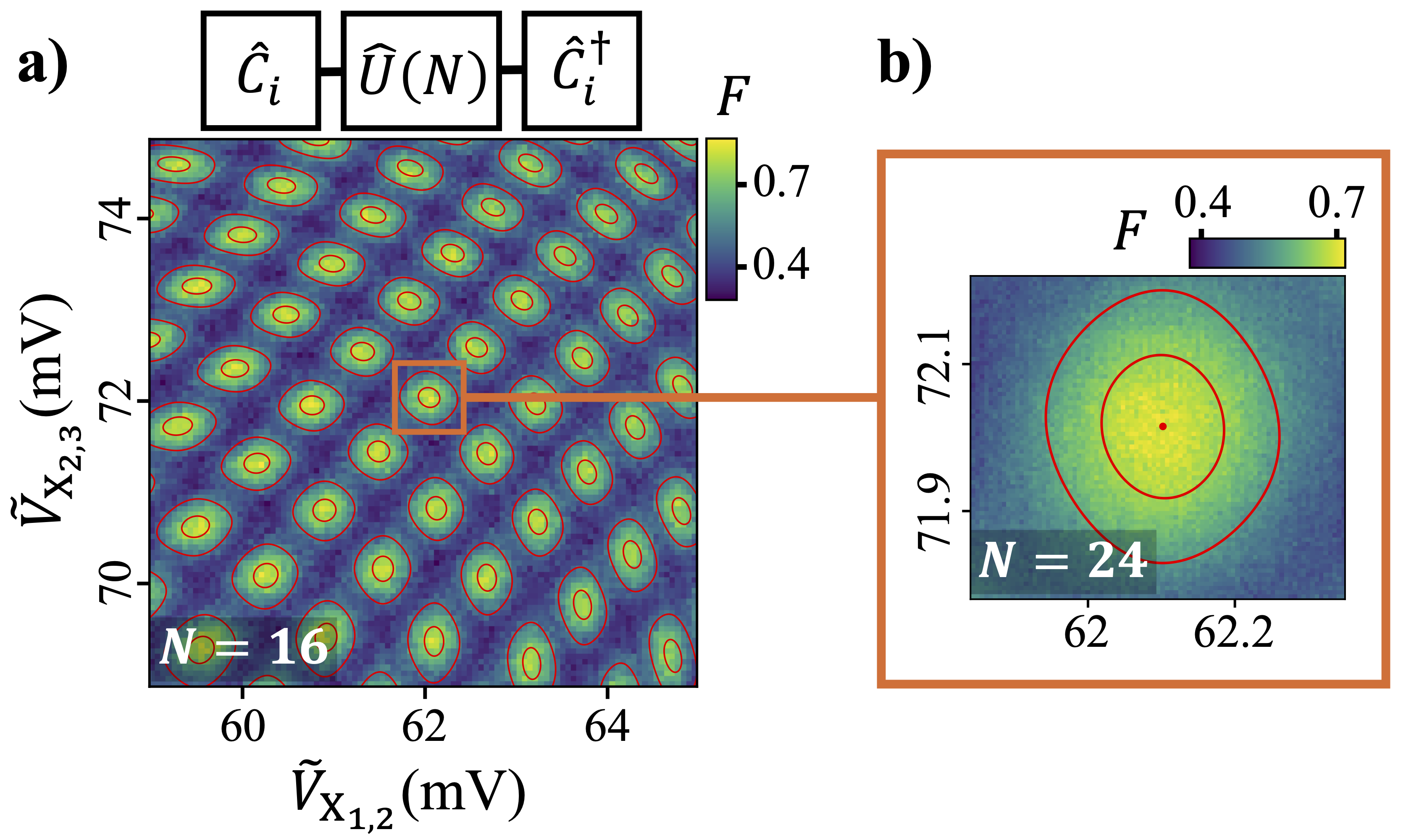}
\caption{\label{fig:figure4} \textbf{2-$J$ gate calibration}.
(a) $F$ obtained by twirling $U$ over the 1-$J$ single-qubit Cliffords as a function of the amplitudes $\tilde{V}_{X_{1,2}}$ and $\tilde{V}_{X_{2,3}}$ of a 2-$J$ exchange pulse. Here, $\hat{U}$ is designed to calibrate a 2-$J$ $\pi$-rotation about the $-\hat{z}$ axis. The calibration is optimized by choosing the pulse amplitudes that maximize the central peak of $F$ (enclosed by the orange square). (b) Zoom-in of the optimal peak with $N = 24$. The red contour lines are obtained by fitting the data to determine the peak location (see Supplementary Information).}
\end{figure}

A specific rotation $\hat{R}_{\varphi}(\theta)$ can be calibrated by sweeping over exchange gate pulse amplitudes to produce a two-dimensional plot of $F$. An example sweep is shown in Fig.~\ref{fig:figure4} for a 2-$J$ $\pi$-rotation about the $-\hat{z}$-axis. Here a pre-calibrated 2-$J$ $\pi$-rotation about the $\hat{x}$-axis is used for $\hat{R}_{\eta}(\chi)$ in the construction of $\hat{U}_{\textrm{ax}}$. $F$ varies periodically with the exchange gate voltages, resulting in a series of interference peaks, with the frequency (and thus sensitivity) increasing with $N$. Only the central peak corresponds to the optimal calibration condition $\varphi=\varphi^*$ and $\theta=\theta^*$, which we track by performing successive sweeps with increasing values of $N=1,2,4,...$, continuing until reductions in the signal-to-noise ratio prevents scaling beyond $N$ = 24.

Using this calibration procedure, we tune nine distinct 2-$J$ exchange pulses corresponding to rotations of $\pi/2$, $\pi$, and $3\pi/2$ about each of the axes: $-\hat{z}$, $-\hat{x}$, and $\hat{x}$. The locations of these pulses in bias space are indicated by the colored markers in Fig.~\ref{fig:figure2}. To evaluate the performance of each pulse, we perform interleaved BRB, which involves interleaving a calibrated 2-$J$ rotation into BRB sequences constructed from 1-$J$ exchange pulses \cite{Andrews2019}.  Details on the calibration procedure and interleaved BRB measurements are given in the Supplementary Information.

\begin{figure}
\centering
\includegraphics[width=0.475\textwidth]{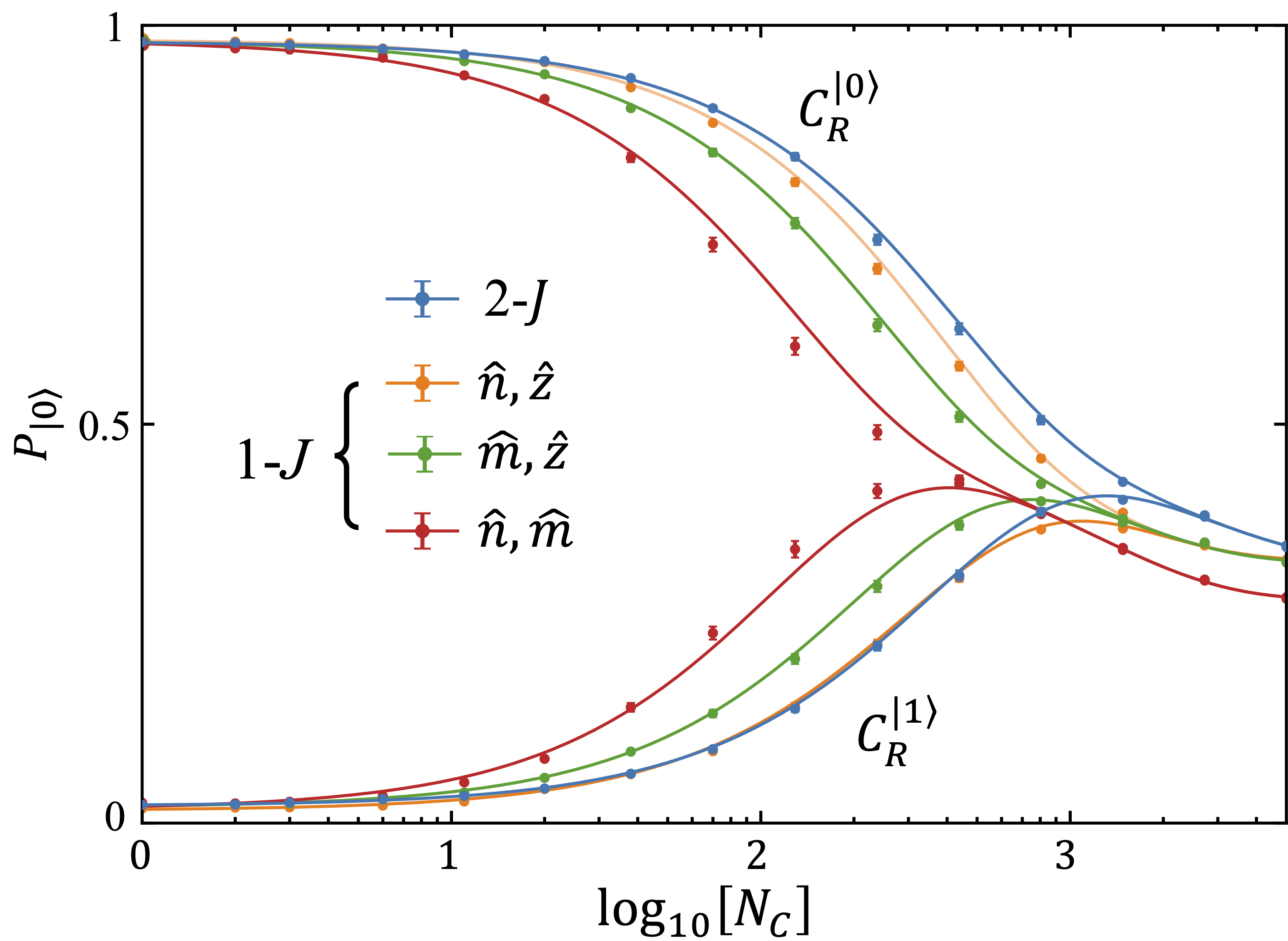}
\caption{\label{fig:figure5} \textbf{Blind randomized benchmarking}. The blue curve shows the results of BRB performed using a Clifford gate set constructed from 2-$J$ exchange pulses. The average single qubit Clifford gate fidelity is $F_{C1}$ = 99.86\% and the average leakage rate is 0.015\% per Clifford. For comparison, we also perform standard 1-$J$ BRB for all pairwise combinations of 1-$J$ exchange axes (blue, green and red curves). In general, 2-$J$ BRB outperforms 1-$J$ BRB, achieving both a higher average Clifford gate fidelity and a lower leakage rate. Error bars indicate the standard error of the mean probability from 250 sequence repetitions.
}
\end{figure}

To quantitatively assess the AEON qubit performance, we execute BRB on 2-$J$ single-qubit Clifford gates constructed from the best performing 2-$J$ pulses~\cite{Andrews2019}.  In Fig.~\ref{fig:figure5} we compare 2-$J$ gate fidelities to the BRB performance of standard 1-$J$ EO single-qubit gates in the same device. Our measurements yield a 2-$J$ single qubit Clifford gate fidelity $F_{C1}$ = 99.86\%, slightly surpassing the best 1-$J$ BRB fidelity of 99.84\% obtained from rotations about the $\hat{n}$ and $\hat{z}$ axes. Additionally, the 2-$J$ leakage error per Clifford, 0.015\%, is approximately half that of the average 1-$J$ leakage error per Clifford, 0.029\%. This reduction is consistent with the leakage-protected nature of the 2-$J$ gates, as each Clifford gate spends roughly half of its duration at idle where exchange is negligible.

Although 2-$J$ BRB outperforms 1-$J$ BRB, this advantage largely stems from shorter gate depths: each 2-$J$ Clifford gate requires an average of 1.9 exchange pulses, compared to 2.7 pulses for 1-$J$ Clifford gates. When accounting for this difference, the total 2-$J$ error per exchange pulse ($\varepsilon_{pp} \approx 0.076\%$) is actually higher than that of the best performing 1-$J$ gates ($\varepsilon_{pp} \approx 0.060\%$), which is surprising given the significantly higher values of $N_{\textrm{osc}}$ observed for 2-$J$ exchange. This discrepancy likely arises from the non-commutativity of 2-$J$ exchange (breakdowns in the approximation of Eq.~\ref{eq:qubit rotation}) due to differences in the transient exchange response to the two simultaneous voltage pulses. Nevertheless, our results highlight the benefits of reduced gate depths and may yet provide further dividends when applied to two-qubit operations \cite{heinz2024}. 

\section{Conclusion}

We have demonstrated high-fidelity quantum control of an AEON qubit that is operated using simultaneous 2-$J$ exchange pulses. The performance of the AEON qubit was validated through BRB, yielding an average Clifford gate fidelity $F_{C1}$ = 99.86\% and an average error per pulse of 0.076\%. Measurements of 2-$J$ exchange oscillations ($N_{\rm osc})$ consistently outperformed 1-$J$ exchange oscillations, suggesting that pulsing simultaneous exchange is less sensitive to charge noise and that the 2-$J$ gate fidelity is currently limited by differences in the transient response of the two exchange interactions. The remaining error could be mitigated in the future by temporal pulse shaping using optimal control protocols \cite{Glaser2015}.

Our work can be extended in several fascinating directions. First, 2-$J$ exchange offers substantial reductions in gate depths for entangling operations, reducing the number of exchange pulses by well over an order of magnitude. Therefore, demonstrating entanglement of two AEON qubits would be a milestone. Second, for single-qubit operation, 3-$J$ exchange enables the construction of a leakage-protected identity gate \cite{Weinstein2005}, which can further suppress leakage errors during idle. Lastly, simultaneous exchange could be utilized to implement more efficient native $i$-Toffoli gates in arrays of Loss-DiVincenzo single-spin qubits \cite{gullans2019protocol}.

\bibliography{maintext_library}

\begin{thebibliography}{45}%
\makeatletter
\providecommand \@ifxundefined [1]{%
 \@ifx{#1\undefined}
}%
\providecommand \@ifnum [1]{%
 \ifnum #1\expandafter \@firstoftwo
 \else \expandafter \@secondoftwo
 \fi
}%
\providecommand \@ifx [1]{%
 \ifx #1\expandafter \@firstoftwo
 \else \expandafter \@secondoftwo
 \fi
}%
\providecommand \natexlab [1]{#1}%
\providecommand \enquote  [1]{``#1''}%
\providecommand \bibnamefont  [1]{#1}%
\providecommand \bibfnamefont [1]{#1}%
\providecommand \citenamefont [1]{#1}%
\providecommand \href@noop [0]{\@secondoftwo}%
\providecommand \href [0]{\begingroup \@sanitize@url \@href}%
\providecommand \@href[1]{\@@startlink{#1}\@@href}%
\providecommand \@@href[1]{\endgroup#1\@@endlink}%
\providecommand \@sanitize@url [0]{\catcode `\\12\catcode `\$12\catcode `\&12\catcode `\#12\catcode `\^12\catcode `\_12\catcode `\%12\relax}%
\providecommand \@@startlink[1]{}%
\providecommand \@@endlink[0]{}%
\providecommand \url  [0]{\begingroup\@sanitize@url \@url }%
\providecommand \@url [1]{\endgroup\@href {#1}{\urlprefix }}%
\providecommand \urlprefix  [0]{URL }%
\providecommand \Eprint [0]{\href }%
\providecommand \doibase [0]{https://doi.org/}%
\providecommand \selectlanguage [0]{\@gobble}%
\providecommand \bibinfo  [0]{\@secondoftwo}%
\providecommand \bibfield  [0]{\@secondoftwo}%
\providecommand \translation [1]{[#1]}%
\providecommand \BibitemOpen [0]{}%
\providecommand \bibitemStop [0]{}%
\providecommand \bibitemNoStop [0]{.\EOS\space}%
\providecommand \EOS [0]{\spacefactor3000\relax}%
\providecommand \BibitemShut  [1]{\csname bibitem#1\endcsname}%
\let\auto@bib@innerbib\@empty
\bibitem [{\citenamefont {Burkard}\ \emph {et~al.}(2023)\citenamefont {Burkard}, \citenamefont {Ladd}, \citenamefont {Pan}, \citenamefont {Nichol},\ and\ \citenamefont {Petta}}]{burkard2023}%
  \BibitemOpen
  \bibfield  {author} {\bibinfo {author} {\bibfnamefont {G.}~\bibnamefont {Burkard}}, \bibinfo {author} {\bibfnamefont {T.~D.}\ \bibnamefont {Ladd}}, \bibinfo {author} {\bibfnamefont {A.}~\bibnamefont {Pan}}, \bibinfo {author} {\bibfnamefont {J.~M.}\ \bibnamefont {Nichol}},\ and\ \bibinfo {author} {\bibfnamefont {J.~R.}\ \bibnamefont {Petta}},\ }\bibfield  {title} {\bibinfo {title} {Semiconductor spin qubits},\ }\href {https://doi.org/10.1103/RevModPhys.95.025003} {\bibfield  {journal} {\bibinfo  {journal} {Rev. Mod. Phys.}\ }\textbf {\bibinfo {volume} {95}},\ \bibinfo {pages} {025003} (\bibinfo {year} {2023})}\BibitemShut {NoStop}%
\bibitem [{\citenamefont {Ansaloni}\ \emph {et~al.}(2020)\citenamefont {Ansaloni}, \citenamefont {Chatterjee}, \citenamefont {Bohuslavskyi}, \citenamefont {Bertrand}, \citenamefont {Hutin}, \citenamefont {Vinet},\ and\ \citenamefont {Kuemmeth}}]{Ansaloni2020}%
  \BibitemOpen
  \bibfield  {author} {\bibinfo {author} {\bibfnamefont {F.}~\bibnamefont {Ansaloni}}, \bibinfo {author} {\bibfnamefont {A.}~\bibnamefont {Chatterjee}}, \bibinfo {author} {\bibfnamefont {H.}~\bibnamefont {Bohuslavskyi}}, \bibinfo {author} {\bibfnamefont {B.}~\bibnamefont {Bertrand}}, \bibinfo {author} {\bibfnamefont {L.}~\bibnamefont {Hutin}}, \bibinfo {author} {\bibfnamefont {M.}~\bibnamefont {Vinet}},\ and\ \bibinfo {author} {\bibfnamefont {F.}~\bibnamefont {Kuemmeth}},\ }\bibfield  {title} {\bibinfo {title} {Single-electron operations in a foundry-fabricated array of quantum dots},\ }\href {https://doi.org/10.1038/s41467-020-20280-3} {\bibfield  {journal} {\bibinfo  {journal} {Nat. Commun.}\ }\textbf {\bibinfo {volume} {11}},\ \bibinfo {pages} {6399} (\bibinfo {year} {2020})}\BibitemShut {NoStop}%
\bibitem [{\citenamefont {Ha}\ \emph {et~al.}(2022)\citenamefont {Ha}, \citenamefont {Ha}, \citenamefont {Choi}, \citenamefont {Tang}, \citenamefont {Schmitz}, \citenamefont {Levendorf}, \citenamefont {Lee}, \citenamefont {Chappell}, \citenamefont {Adams}, \citenamefont {Hulbert}, \citenamefont {Acuna}, \citenamefont {Noah}, \citenamefont {Matten}, \citenamefont {Jura}, \citenamefont {Wright}, \citenamefont {Rakher},\ and\ \citenamefont {Borselli}}]{ha2022}%
  \BibitemOpen
  \bibfield  {author} {\bibinfo {author} {\bibfnamefont {W.}~\bibnamefont {Ha}}, \bibinfo {author} {\bibfnamefont {S.~D.}\ \bibnamefont {Ha}}, \bibinfo {author} {\bibfnamefont {M.~D.}\ \bibnamefont {Choi}}, \bibinfo {author} {\bibfnamefont {Y.}~\bibnamefont {Tang}}, \bibinfo {author} {\bibfnamefont {A.~E.}\ \bibnamefont {Schmitz}}, \bibinfo {author} {\bibfnamefont {M.~P.}\ \bibnamefont {Levendorf}}, \bibinfo {author} {\bibfnamefont {K.}~\bibnamefont {Lee}}, \bibinfo {author} {\bibfnamefont {J.~M.}\ \bibnamefont {Chappell}}, \bibinfo {author} {\bibfnamefont {T.~S.}\ \bibnamefont {Adams}}, \bibinfo {author} {\bibfnamefont {D.~R.}\ \bibnamefont {Hulbert}}, \bibinfo {author} {\bibfnamefont {E.}~\bibnamefont {Acuna}}, \bibinfo {author} {\bibfnamefont {R.~S.}\ \bibnamefont {Noah}}, \bibinfo {author} {\bibfnamefont {J.~W.}\ \bibnamefont {Matten}}, \bibinfo {author} {\bibfnamefont {M.~P.}\ \bibnamefont {Jura}}, \bibinfo {author} {\bibfnamefont {J.~A.}\ \bibnamefont {Wright}}, \bibinfo {author} {\bibfnamefont {M.~T.}\ \bibnamefont {Rakher}},\ and\ \bibinfo {author} {\bibfnamefont {M.~G.}\ \bibnamefont {Borselli}},\ }\bibfield  {title} {\bibinfo {title} {{A Flexible Design Platform for Si/SiGe Exchange-Only Qubits with Low Disorder}},\ }\href {https://doi.org/10.1021/acs.nanolett.1c03026} {\bibfield  {journal} {\bibinfo  {journal} {Nano Lett.}\ }\textbf {\bibinfo {volume} {22}},\ \bibinfo {pages} {1443} (\bibinfo {year} {2022})}\BibitemShut {NoStop}%
\bibitem [{\citenamefont {Zwerver}\ \emph {et~al.}(2022)\citenamefont {Zwerver}, \citenamefont {Kr{\"a}henmann}, \citenamefont {Watson}, \citenamefont {Lampert}, \citenamefont {George}, \citenamefont {Pillarisetty}, \citenamefont {Bojarski}, \citenamefont {Amin}, \citenamefont {Amitonov}, \citenamefont {Boter}, \citenamefont {Caudillo}, \citenamefont {Correas-Serrano}, \citenamefont {Dehollain}, \citenamefont {Droulers}, \citenamefont {Henry}, \citenamefont {Kotlyar}, \citenamefont {Lodari}, \citenamefont {L{\"u}thi}, \citenamefont {Michalak}, \citenamefont {Mueller}, \citenamefont {Neyens}, \citenamefont {Roberts}, \citenamefont {Samkharadze}, \citenamefont {Zheng}, \citenamefont {Zietz}, \citenamefont {Scappucci}, \citenamefont {Veldhorst}, \citenamefont {Vandersypen},\ and\ \citenamefont {Clarke}}]{Zwerver2022}%
  \BibitemOpen
  \bibfield  {author} {\bibinfo {author} {\bibfnamefont {A.~M.~J.}\ \bibnamefont {Zwerver}}, \bibinfo {author} {\bibfnamefont {T.}~\bibnamefont {Kr{\"a}henmann}}, \bibinfo {author} {\bibfnamefont {T.~F.}\ \bibnamefont {Watson}}, \bibinfo {author} {\bibfnamefont {L.}~\bibnamefont {Lampert}}, \bibinfo {author} {\bibfnamefont {H.~C.}\ \bibnamefont {George}}, \bibinfo {author} {\bibfnamefont {R.}~\bibnamefont {Pillarisetty}}, \bibinfo {author} {\bibfnamefont {S.~A.}\ \bibnamefont {Bojarski}}, \bibinfo {author} {\bibfnamefont {P.}~\bibnamefont {Amin}}, \bibinfo {author} {\bibfnamefont {S.~V.}\ \bibnamefont {Amitonov}}, \bibinfo {author} {\bibfnamefont {J.~M.}\ \bibnamefont {Boter}}, \bibinfo {author} {\bibfnamefont {R.}~\bibnamefont {Caudillo}}, \bibinfo {author} {\bibfnamefont {D.}~\bibnamefont {Correas-Serrano}}, \bibinfo {author} {\bibfnamefont {J.~P.}\ \bibnamefont {Dehollain}}, \bibinfo {author} {\bibfnamefont {G.}~\bibnamefont {Droulers}}, \bibinfo {author} {\bibfnamefont {E.~M.}\ \bibnamefont {Henry}}, \bibinfo {author} {\bibfnamefont {R.}~\bibnamefont {Kotlyar}}, \bibinfo {author} {\bibfnamefont {M.}~\bibnamefont {Lodari}}, \bibinfo {author} {\bibfnamefont {F.}~\bibnamefont {L{\"u}thi}}, \bibinfo {author} {\bibfnamefont {D.~J.}\ \bibnamefont {Michalak}}, \bibinfo {author} {\bibfnamefont {B.~K.}\ \bibnamefont {Mueller}}, \bibinfo {author} {\bibfnamefont {S.}~\bibnamefont {Neyens}}, \bibinfo {author} {\bibfnamefont {J.}~\bibnamefont {Roberts}}, \bibinfo {author} {\bibfnamefont {N.}~\bibnamefont {Samkharadze}}, \bibinfo {author} {\bibfnamefont {G.}~\bibnamefont {Zheng}}, \bibinfo {author} {\bibfnamefont {O.~K.}\ \bibnamefont {Zietz}}, \bibinfo {author} {\bibfnamefont {G.}~\bibnamefont {Scappucci}}, \bibinfo {author} {\bibfnamefont {M.}~\bibnamefont {Veldhorst}}, \bibinfo {author} {\bibfnamefont {L.~M.~K.}\ \bibnamefont {Vandersypen}},\ and\ \bibinfo {author} {\bibfnamefont {J.~S.}\ \bibnamefont {Clarke}},\ }\bibfield  {title} {\bibinfo {title} {Qubits made by advanced semiconductor manufacturing},\ }\href {https://doi.org/10.1038/s41928-022-00727-9} {\bibfield  {journal} {\bibinfo  {journal} {Nat. Electron.}\ }\textbf {\bibinfo {volume} {5}},\ \bibinfo {pages} {184} (\bibinfo {year} {2022})}\BibitemShut {NoStop}%
\bibitem [{\citenamefont {Neyens}\ \emph {et~al.}(2024)\citenamefont {Neyens}, \citenamefont {Zietz}, \citenamefont {Watson}, \citenamefont {Luthi}, \citenamefont {Nethwewala}, \citenamefont {George}, \citenamefont {Henry}, \citenamefont {Islam}, \citenamefont {Wagner}, \citenamefont {Borjans}, \citenamefont {Connors}, \citenamefont {Corrigan}, \citenamefont {Curry}, \citenamefont {Keith}, \citenamefont {Kotlyar}, \citenamefont {Lampert}, \citenamefont {M{\k{a}}dzik}, \citenamefont {Millard}, \citenamefont {Mohiyaddin}, \citenamefont {Pellerano}, \citenamefont {Pillarisetty}, \citenamefont {Ramsey}, \citenamefont {Savytskyy}, \citenamefont {Schaal}, \citenamefont {Zheng}, \citenamefont {Ziegler}, \citenamefont {Bishop}, \citenamefont {Bojarski}, \citenamefont {Roberts},\ and\ \citenamefont {Clarke}}]{Neyens2024}%
  \BibitemOpen
  \bibfield  {author} {\bibinfo {author} {\bibfnamefont {S.}~\bibnamefont {Neyens}}, \bibinfo {author} {\bibfnamefont {O.~K.}\ \bibnamefont {Zietz}}, \bibinfo {author} {\bibfnamefont {T.~F.}\ \bibnamefont {Watson}}, \bibinfo {author} {\bibfnamefont {F.}~\bibnamefont {Luthi}}, \bibinfo {author} {\bibfnamefont {A.}~\bibnamefont {Nethwewala}}, \bibinfo {author} {\bibfnamefont {H.~C.}\ \bibnamefont {George}}, \bibinfo {author} {\bibfnamefont {E.}~\bibnamefont {Henry}}, \bibinfo {author} {\bibfnamefont {M.}~\bibnamefont {Islam}}, \bibinfo {author} {\bibfnamefont {A.~J.}\ \bibnamefont {Wagner}}, \bibinfo {author} {\bibfnamefont {F.}~\bibnamefont {Borjans}}, \bibinfo {author} {\bibfnamefont {E.~J.}\ \bibnamefont {Connors}}, \bibinfo {author} {\bibfnamefont {J.}~\bibnamefont {Corrigan}}, \bibinfo {author} {\bibfnamefont {M.~J.}\ \bibnamefont {Curry}}, \bibinfo {author} {\bibfnamefont {D.}~\bibnamefont {Keith}}, \bibinfo {author} {\bibfnamefont {R.}~\bibnamefont {Kotlyar}}, \bibinfo {author} {\bibfnamefont {L.~F.}\ \bibnamefont {Lampert}}, \bibinfo {author} {\bibfnamefont {M.~T.}\ \bibnamefont {M{\k{a}}dzik}}, \bibinfo {author} {\bibfnamefont {K.}~\bibnamefont {Millard}}, \bibinfo {author} {\bibfnamefont {F.~A.}\ \bibnamefont {Mohiyaddin}}, \bibinfo {author} {\bibfnamefont {S.}~\bibnamefont {Pellerano}}, \bibinfo {author} {\bibfnamefont {R.}~\bibnamefont {Pillarisetty}}, \bibinfo {author} {\bibfnamefont {M.}~\bibnamefont {Ramsey}}, \bibinfo {author} {\bibfnamefont {R.}~\bibnamefont {Savytskyy}}, \bibinfo {author} {\bibfnamefont {S.}~\bibnamefont {Schaal}}, \bibinfo {author} {\bibfnamefont {G.}~\bibnamefont {Zheng}}, \bibinfo {author} {\bibfnamefont {J.}~\bibnamefont {Ziegler}}, \bibinfo {author} {\bibfnamefont {N.~C.}\ \bibnamefont {Bishop}}, \bibinfo {author} {\bibfnamefont {S.}~\bibnamefont {Bojarski}}, \bibinfo {author} {\bibfnamefont {J.}~\bibnamefont {Roberts}},\ and\ \bibinfo {author} {\bibfnamefont {J.~S.}\ \bibnamefont {Clarke}},\ }\bibfield  {title} {\bibinfo {title} {Probing single electrons across 300-mm spin qubit wafers},\ }\href {https://doi.org/10.1038/s41586-024-07275-6} {\bibfield  {journal} {\bibinfo  {journal} {Nature}\ }\textbf {\bibinfo {volume} {629}},\ \bibinfo {pages} {80} (\bibinfo {year} {2024})}\BibitemShut {NoStop}%
\bibitem [{\citenamefont {Petit}\ \emph {et~al.}(2020)\citenamefont {Petit}, \citenamefont {Eenink}, \citenamefont {Russ}, \citenamefont {Lawrie}, \citenamefont {Hendrickx}, \citenamefont {Philips}, \citenamefont {Clarke}, \citenamefont {Vandersypen},\ and\ \citenamefont {Veldhorst}}]{Petit2020}%
  \BibitemOpen
  \bibfield  {author} {\bibinfo {author} {\bibfnamefont {L.}~\bibnamefont {Petit}}, \bibinfo {author} {\bibfnamefont {H.~G.~J.}\ \bibnamefont {Eenink}}, \bibinfo {author} {\bibfnamefont {M.}~\bibnamefont {Russ}}, \bibinfo {author} {\bibfnamefont {W.~I.~L.}\ \bibnamefont {Lawrie}}, \bibinfo {author} {\bibfnamefont {N.~W.}\ \bibnamefont {Hendrickx}}, \bibinfo {author} {\bibfnamefont {S.~G.~J.}\ \bibnamefont {Philips}}, \bibinfo {author} {\bibfnamefont {J.~S.}\ \bibnamefont {Clarke}}, \bibinfo {author} {\bibfnamefont {L.~M.~K.}\ \bibnamefont {Vandersypen}},\ and\ \bibinfo {author} {\bibfnamefont {M.}~\bibnamefont {Veldhorst}},\ }\bibfield  {title} {\bibinfo {title} {Universal quantum logic in hot silicon qubits},\ }\href {https://doi.org/10.1038/s41586-020-2170-7} {\bibfield  {journal} {\bibinfo  {journal} {Nature}\ }\textbf {\bibinfo {volume} {580}},\ \bibinfo {pages} {355} (\bibinfo {year} {2020})}\BibitemShut {NoStop}%
\bibitem [{\citenamefont {Yang}\ \emph {et~al.}(2020)\citenamefont {Yang}, \citenamefont {Leon}, \citenamefont {Hwang}, \citenamefont {Saraiva}, \citenamefont {Tanttu}, \citenamefont {Huang}, \citenamefont {Camirand~Lemyre}, \citenamefont {Chan}, \citenamefont {Tan}, \citenamefont {Hudson}, \citenamefont {Itoh}, \citenamefont {Morello}, \citenamefont {Pioro-Ladri{\`e}re}, \citenamefont {Laucht},\ and\ \citenamefont {Dzurak}}]{Yang2020}%
  \BibitemOpen
  \bibfield  {author} {\bibinfo {author} {\bibfnamefont {C.~H.}\ \bibnamefont {Yang}}, \bibinfo {author} {\bibfnamefont {R.~C.~C.}\ \bibnamefont {Leon}}, \bibinfo {author} {\bibfnamefont {J.~C.~C.}\ \bibnamefont {Hwang}}, \bibinfo {author} {\bibfnamefont {A.}~\bibnamefont {Saraiva}}, \bibinfo {author} {\bibfnamefont {T.}~\bibnamefont {Tanttu}}, \bibinfo {author} {\bibfnamefont {W.}~\bibnamefont {Huang}}, \bibinfo {author} {\bibfnamefont {J.}~\bibnamefont {Camirand~Lemyre}}, \bibinfo {author} {\bibfnamefont {K.~W.}\ \bibnamefont {Chan}}, \bibinfo {author} {\bibfnamefont {K.~Y.}\ \bibnamefont {Tan}}, \bibinfo {author} {\bibfnamefont {F.~E.}\ \bibnamefont {Hudson}}, \bibinfo {author} {\bibfnamefont {K.~M.}\ \bibnamefont {Itoh}}, \bibinfo {author} {\bibfnamefont {A.}~\bibnamefont {Morello}}, \bibinfo {author} {\bibfnamefont {M.}~\bibnamefont {Pioro-Ladri{\`e}re}}, \bibinfo {author} {\bibfnamefont {A.}~\bibnamefont {Laucht}},\ and\ \bibinfo {author} {\bibfnamefont {A.~S.}\ \bibnamefont {Dzurak}},\ }\bibfield  {title} {\bibinfo {title} {Operation of a silicon quantum processor unit cell above one kelvin},\ }\href {https://doi.org/10.1038/s41586-020-2171-6} {\bibfield  {journal} {\bibinfo  {journal} {Nature}\ }\textbf {\bibinfo {volume} {580}},\ \bibinfo {pages} {350} (\bibinfo {year} {2020})}\BibitemShut {NoStop}%
\bibitem [{\citenamefont {Veldhorst}\ \emph {et~al.}(2014)\citenamefont {Veldhorst}, \citenamefont {Hwang}, \citenamefont {Yang}, \citenamefont {Leenstra}, \citenamefont {de~Ronde}, \citenamefont {Dehollain}, \citenamefont {Muhonen}, \citenamefont {Hudson}, \citenamefont {Itoh}, \citenamefont {Morello},\ and\ \citenamefont {Dzurak}}]{Veldhorst2014}%
  \BibitemOpen
  \bibfield  {author} {\bibinfo {author} {\bibfnamefont {M.}~\bibnamefont {Veldhorst}}, \bibinfo {author} {\bibfnamefont {J.~C.~C.}\ \bibnamefont {Hwang}}, \bibinfo {author} {\bibfnamefont {C.~H.}\ \bibnamefont {Yang}}, \bibinfo {author} {\bibfnamefont {A.~W.}\ \bibnamefont {Leenstra}}, \bibinfo {author} {\bibfnamefont {B.}~\bibnamefont {de~Ronde}}, \bibinfo {author} {\bibfnamefont {J.~P.}\ \bibnamefont {Dehollain}}, \bibinfo {author} {\bibfnamefont {J.~T.}\ \bibnamefont {Muhonen}}, \bibinfo {author} {\bibfnamefont {F.~E.}\ \bibnamefont {Hudson}}, \bibinfo {author} {\bibfnamefont {K.~M.}\ \bibnamefont {Itoh}}, \bibinfo {author} {\bibfnamefont {A.}~\bibnamefont {Morello}},\ and\ \bibinfo {author} {\bibfnamefont {A.~S.}\ \bibnamefont {Dzurak}},\ }\bibfield  {title} {\bibinfo {title} {An addressable quantum dot qubit with fault-tolerant control-fidelity},\ }\href {https://doi.org/10.1038/nnano.2014.216} {\bibfield  {journal} {\bibinfo  {journal} {Nat. Nanotechnol.}\ }\textbf {\bibinfo {volume} {9}},\ \bibinfo {pages} {981} (\bibinfo {year} {2014})}\BibitemShut {NoStop}%
\bibitem [{\citenamefont {Muhonen}\ \emph {et~al.}(2014)\citenamefont {Muhonen}, \citenamefont {Dehollain}, \citenamefont {Laucht}, \citenamefont {Hudson}, \citenamefont {Kalra}, \citenamefont {Sekiguchi}, \citenamefont {Itoh}, \citenamefont {Jamieson}, \citenamefont {McCallum}, \citenamefont {Dzurak},\ and\ \citenamefont {Morello}}]{Muhonen2014}%
  \BibitemOpen
  \bibfield  {author} {\bibinfo {author} {\bibfnamefont {J.~T.}\ \bibnamefont {Muhonen}}, \bibinfo {author} {\bibfnamefont {J.~P.}\ \bibnamefont {Dehollain}}, \bibinfo {author} {\bibfnamefont {A.}~\bibnamefont {Laucht}}, \bibinfo {author} {\bibfnamefont {F.~E.}\ \bibnamefont {Hudson}}, \bibinfo {author} {\bibfnamefont {R.}~\bibnamefont {Kalra}}, \bibinfo {author} {\bibfnamefont {T.}~\bibnamefont {Sekiguchi}}, \bibinfo {author} {\bibfnamefont {K.~M.}\ \bibnamefont {Itoh}}, \bibinfo {author} {\bibfnamefont {D.~N.}\ \bibnamefont {Jamieson}}, \bibinfo {author} {\bibfnamefont {J.~C.}\ \bibnamefont {McCallum}}, \bibinfo {author} {\bibfnamefont {A.~S.}\ \bibnamefont {Dzurak}},\ and\ \bibinfo {author} {\bibfnamefont {A.}~\bibnamefont {Morello}},\ }\bibfield  {title} {\bibinfo {title} {Storing quantum information for 30 seconds in a nanoelectronic device},\ }\href {https://doi.org/10.1038/nnano.2014.211} {\bibfield  {journal} {\bibinfo  {journal} {Nat. Nanotechnol.}\ }\textbf {\bibinfo {volume} {9}},\ \bibinfo {pages} {986} (\bibinfo {year} {2014})}\BibitemShut {NoStop}%
\bibitem [{\citenamefont {Loss}\ and\ \citenamefont {DiVincenzo}(1998)}]{loss1998}%
  \BibitemOpen
  \bibfield  {author} {\bibinfo {author} {\bibfnamefont {D.}~\bibnamefont {Loss}}\ and\ \bibinfo {author} {\bibfnamefont {D.~P.}\ \bibnamefont {DiVincenzo}},\ }\bibfield  {title} {\bibinfo {title} {Quantum computation with quantum dots},\ }\href {https://doi.org/10.1103/PhysRevA.57.120} {\bibfield  {journal} {\bibinfo  {journal} {Phys. Rev. A}\ }\textbf {\bibinfo {volume} {57}},\ \bibinfo {pages} {120} (\bibinfo {year} {1998})}\BibitemShut {NoStop}%
\bibitem [{\citenamefont {Petta}\ \emph {et~al.}(2005)\citenamefont {Petta}, \citenamefont {Johnson}, \citenamefont {Taylor}, \citenamefont {Laird}, \citenamefont {Yacoby}, \citenamefont {Lukin}, \citenamefont {Marcus}, \citenamefont {Hanson},\ and\ \citenamefont {Gossard}}]{petta2005}%
  \BibitemOpen
  \bibfield  {author} {\bibinfo {author} {\bibfnamefont {J.~R.}\ \bibnamefont {Petta}}, \bibinfo {author} {\bibfnamefont {A.~C.}\ \bibnamefont {Johnson}}, \bibinfo {author} {\bibfnamefont {J.~M.}\ \bibnamefont {Taylor}}, \bibinfo {author} {\bibfnamefont {E.~A.}\ \bibnamefont {Laird}}, \bibinfo {author} {\bibfnamefont {A.}~\bibnamefont {Yacoby}}, \bibinfo {author} {\bibfnamefont {M.~D.}\ \bibnamefont {Lukin}}, \bibinfo {author} {\bibfnamefont {C.~M.}\ \bibnamefont {Marcus}}, \bibinfo {author} {\bibfnamefont {M.~P.}\ \bibnamefont {Hanson}},\ and\ \bibinfo {author} {\bibfnamefont {A.~C.}\ \bibnamefont {Gossard}},\ }\bibfield  {title} {\bibinfo {title} {Coherent manipulation of coupled electron spins in semiconductor quantum dots},\ }\href {https://doi.org/10.1126/science.1116955} {\bibfield  {journal} {\bibinfo  {journal} {Science}\ }\textbf {\bibinfo {volume} {309}},\ \bibinfo {pages} {2180} (\bibinfo {year} {2005})}\BibitemShut {NoStop}%
\bibitem [{\citenamefont {Tokura}\ \emph {et~al.}(2006)\citenamefont {Tokura}, \citenamefont {van~der Wiel}, \citenamefont {Obata},\ and\ \citenamefont {Tarucha}}]{tokura2006}%
  \BibitemOpen
  \bibfield  {author} {\bibinfo {author} {\bibfnamefont {Y.}~\bibnamefont {Tokura}}, \bibinfo {author} {\bibfnamefont {W.~G.}\ \bibnamefont {van~der Wiel}}, \bibinfo {author} {\bibfnamefont {T.}~\bibnamefont {Obata}},\ and\ \bibinfo {author} {\bibfnamefont {S.}~\bibnamefont {Tarucha}},\ }\bibfield  {title} {\bibinfo {title} {{Coherent Single Electron Spin Control in a Slanting Zeeman Field}},\ }\href {https://doi.org/10.1103/PhysRevLett.96.047202} {\bibfield  {journal} {\bibinfo  {journal} {Phys. Rev. Lett.}\ }\textbf {\bibinfo {volume} {96}},\ \bibinfo {pages} {047202} (\bibinfo {year} {2006})}\BibitemShut {NoStop}%
\bibitem [{\citenamefont {Pioro-Ladri{\`e}re}\ \emph {et~al.}(2008)\citenamefont {Pioro-Ladri{\`e}re}, \citenamefont {Obata}, \citenamefont {Tokura}, \citenamefont {Shin}, \citenamefont {Kubo}, \citenamefont {Yoshida}, \citenamefont {Taniyama},\ and\ \citenamefont {Tarucha}}]{Pioro2008}%
  \BibitemOpen
  \bibfield  {author} {\bibinfo {author} {\bibfnamefont {M.}~\bibnamefont {Pioro-Ladri{\`e}re}}, \bibinfo {author} {\bibfnamefont {T.}~\bibnamefont {Obata}}, \bibinfo {author} {\bibfnamefont {Y.}~\bibnamefont {Tokura}}, \bibinfo {author} {\bibfnamefont {Y.-S.}\ \bibnamefont {Shin}}, \bibinfo {author} {\bibfnamefont {T.}~\bibnamefont {Kubo}}, \bibinfo {author} {\bibfnamefont {K.}~\bibnamefont {Yoshida}}, \bibinfo {author} {\bibfnamefont {T.}~\bibnamefont {Taniyama}},\ and\ \bibinfo {author} {\bibfnamefont {S.}~\bibnamefont {Tarucha}},\ }\bibfield  {title} {\bibinfo {title} {{Electrically driven single-electron spin resonance in a slanting Zeeman field}},\ }\href {https://doi.org/10.1038/nphys1053} {\bibfield  {journal} {\bibinfo  {journal} {Nat. Phys.}\ }\textbf {\bibinfo {volume} {4}},\ \bibinfo {pages} {776} (\bibinfo {year} {2008})}\BibitemShut {NoStop}%
\bibitem [{\citenamefont {Watzinger}\ \emph {et~al.}(2018)\citenamefont {Watzinger}, \citenamefont {Kuku{\v{c}}ka}, \citenamefont {Vuku{\v{s}}i{\'{c}}}, \citenamefont {Gao}, \citenamefont {Wang}, \citenamefont {Sch{\"a}ffler}, \citenamefont {Zhang},\ and\ \citenamefont {Katsaros}}]{Watzinger2018}%
  \BibitemOpen
  \bibfield  {author} {\bibinfo {author} {\bibfnamefont {H.}~\bibnamefont {Watzinger}}, \bibinfo {author} {\bibfnamefont {J.}~\bibnamefont {Kuku{\v{c}}ka}}, \bibinfo {author} {\bibfnamefont {L.}~\bibnamefont {Vuku{\v{s}}i{\'{c}}}}, \bibinfo {author} {\bibfnamefont {F.}~\bibnamefont {Gao}}, \bibinfo {author} {\bibfnamefont {T.}~\bibnamefont {Wang}}, \bibinfo {author} {\bibfnamefont {F.}~\bibnamefont {Sch{\"a}ffler}}, \bibinfo {author} {\bibfnamefont {J.-J.}\ \bibnamefont {Zhang}},\ and\ \bibinfo {author} {\bibfnamefont {G.}~\bibnamefont {Katsaros}},\ }\bibfield  {title} {\bibinfo {title} {A germanium hole spin qubit},\ }\href {https://doi.org/10.1038/s41467-018-06418-4} {\bibfield  {journal} {\bibinfo  {journal} {Nat. Commun.}\ }\textbf {\bibinfo {volume} {9}},\ \bibinfo {pages} {3902} (\bibinfo {year} {2018})}\BibitemShut {NoStop}%
\bibitem [{\citenamefont {DiVincenzo}\ \emph {et~al.}(2000)\citenamefont {DiVincenzo}, \citenamefont {Bacon}, \citenamefont {Kempe}, \citenamefont {Burkard},\ and\ \citenamefont {Whaley}}]{DiVincenzo2000}%
  \BibitemOpen
  \bibfield  {author} {\bibinfo {author} {\bibfnamefont {D.~P.}\ \bibnamefont {DiVincenzo}}, \bibinfo {author} {\bibfnamefont {D.}~\bibnamefont {Bacon}}, \bibinfo {author} {\bibfnamefont {J.}~\bibnamefont {Kempe}}, \bibinfo {author} {\bibfnamefont {G.}~\bibnamefont {Burkard}},\ and\ \bibinfo {author} {\bibfnamefont {K.~B.}\ \bibnamefont {Whaley}},\ }\bibfield  {title} {\bibinfo {title} {Universal quantum computation with the exchange interaction},\ }\href {https://doi.org/10.1038/35042541} {\bibfield  {journal} {\bibinfo  {journal} {Nature}\ }\textbf {\bibinfo {volume} {408}},\ \bibinfo {pages} {339} (\bibinfo {year} {2000})}\BibitemShut {NoStop}%
\bibitem [{\citenamefont {Laird}\ \emph {et~al.}(2010)\citenamefont {Laird}, \citenamefont {Taylor}, \citenamefont {DiVincenzo}, \citenamefont {Marcus}, \citenamefont {Hanson},\ and\ \citenamefont {Gossard}}]{laird2010}%
  \BibitemOpen
  \bibfield  {author} {\bibinfo {author} {\bibfnamefont {E.~A.}\ \bibnamefont {Laird}}, \bibinfo {author} {\bibfnamefont {J.~M.}\ \bibnamefont {Taylor}}, \bibinfo {author} {\bibfnamefont {D.~P.}\ \bibnamefont {DiVincenzo}}, \bibinfo {author} {\bibfnamefont {C.~M.}\ \bibnamefont {Marcus}}, \bibinfo {author} {\bibfnamefont {M.~P.}\ \bibnamefont {Hanson}},\ and\ \bibinfo {author} {\bibfnamefont {A.~C.}\ \bibnamefont {Gossard}},\ }\bibfield  {title} {\bibinfo {title} {Coherent spin manipulation in an exchange-only qubit},\ }\href {https://doi.org/10.1103/PhysRevB.82.075403} {\bibfield  {journal} {\bibinfo  {journal} {Phys. Rev. B}\ }\textbf {\bibinfo {volume} {82}},\ \bibinfo {pages} {075403} (\bibinfo {year} {2010})}\BibitemShut {NoStop}%
\bibitem [{\citenamefont {Medford}\ \emph {et~al.}(2013)\citenamefont {Medford}, \citenamefont {Beil}, \citenamefont {Taylor}, \citenamefont {Bartlett}, \citenamefont {Doherty}, \citenamefont {Rashba}, \citenamefont {DiVincenzo}, \citenamefont {Lu}, \citenamefont {Gossard},\ and\ \citenamefont {Marcus}}]{Medford2013}%
  \BibitemOpen
  \bibfield  {author} {\bibinfo {author} {\bibfnamefont {J.}~\bibnamefont {Medford}}, \bibinfo {author} {\bibfnamefont {J.}~\bibnamefont {Beil}}, \bibinfo {author} {\bibfnamefont {J.~M.}\ \bibnamefont {Taylor}}, \bibinfo {author} {\bibfnamefont {S.~D.}\ \bibnamefont {Bartlett}}, \bibinfo {author} {\bibfnamefont {A.~C.}\ \bibnamefont {Doherty}}, \bibinfo {author} {\bibfnamefont {E.~I.}\ \bibnamefont {Rashba}}, \bibinfo {author} {\bibfnamefont {D.~P.}\ \bibnamefont {DiVincenzo}}, \bibinfo {author} {\bibfnamefont {H.}~\bibnamefont {Lu}}, \bibinfo {author} {\bibfnamefont {A.~C.}\ \bibnamefont {Gossard}},\ and\ \bibinfo {author} {\bibfnamefont {C.~M.}\ \bibnamefont {Marcus}},\ }\bibfield  {title} {\bibinfo {title} {Self-consistent measurement and state tomography of an exchange-only spin qubit},\ }\href {https://doi.org/10.1038/nnano.2013.168} {\bibfield  {journal} {\bibinfo  {journal} {Nat. Nanotechnol.}\ }\textbf {\bibinfo {volume} {8}},\ \bibinfo {pages} {654} (\bibinfo {year} {2013})}\BibitemShut {NoStop}%
\bibitem [{\citenamefont {Andrews}\ \emph {et~al.}(2019)\citenamefont {Andrews}, \citenamefont {Jones}, \citenamefont {Reed}, \citenamefont {Jones}, \citenamefont {Ha}, \citenamefont {Jura}, \citenamefont {Kerckhoff}, \citenamefont {Levendorf}, \citenamefont {Meenehan}, \citenamefont {Merkel}, \citenamefont {Smith}, \citenamefont {Sun}, \citenamefont {Weinstein}, \citenamefont {Rakher}, \citenamefont {Ladd},\ and\ \citenamefont {Borselli}}]{Andrews2019}%
  \BibitemOpen
  \bibfield  {author} {\bibinfo {author} {\bibfnamefont {R.~W.}\ \bibnamefont {Andrews}}, \bibinfo {author} {\bibfnamefont {C.}~\bibnamefont {Jones}}, \bibinfo {author} {\bibfnamefont {M.~D.}\ \bibnamefont {Reed}}, \bibinfo {author} {\bibfnamefont {A.~M.}\ \bibnamefont {Jones}}, \bibinfo {author} {\bibfnamefont {S.~D.}\ \bibnamefont {Ha}}, \bibinfo {author} {\bibfnamefont {M.~P.}\ \bibnamefont {Jura}}, \bibinfo {author} {\bibfnamefont {J.}~\bibnamefont {Kerckhoff}}, \bibinfo {author} {\bibfnamefont {M.}~\bibnamefont {Levendorf}}, \bibinfo {author} {\bibfnamefont {S.}~\bibnamefont {Meenehan}}, \bibinfo {author} {\bibfnamefont {S.~T.}\ \bibnamefont {Merkel}}, \bibinfo {author} {\bibfnamefont {A.}~\bibnamefont {Smith}}, \bibinfo {author} {\bibfnamefont {B.}~\bibnamefont {Sun}}, \bibinfo {author} {\bibfnamefont {A.~J.}\ \bibnamefont {Weinstein}}, \bibinfo {author} {\bibfnamefont {M.~T.}\ \bibnamefont {Rakher}}, \bibinfo {author} {\bibfnamefont {T.~D.}\ \bibnamefont {Ladd}},\ and\ \bibinfo {author} {\bibfnamefont {M.~G.}\ \bibnamefont {Borselli}},\ }\bibfield  {title} {\bibinfo {title} {{Quantifying error and leakage in an encoded Si/SiGe triple-dot qubit}},\ }\href {https://doi.org/10.1038/s41565-019-0500-4} {\bibfield  {journal} {\bibinfo  {journal} {Nat. Nanotechnol.}\ }\textbf {\bibinfo {volume} {14}},\ \bibinfo {pages} {747} (\bibinfo {year} {2019})}\BibitemShut {NoStop}%
\bibitem [{\citenamefont {Weinstein}\ \emph {et~al.}(2023)\citenamefont {Weinstein}, \citenamefont {Reed}, \citenamefont {Jones}, \citenamefont {Andrews}, \citenamefont {Barnes}, \citenamefont {Blumoff}, \citenamefont {Euliss}, \citenamefont {Eng}, \citenamefont {Fong}, \citenamefont {Ha}, \citenamefont {Hulbert}, \citenamefont {Jackson}, \citenamefont {Jura}, \citenamefont {Keating}, \citenamefont {Kerckhoff}, \citenamefont {Kiselev}, \citenamefont {Matten}, \citenamefont {Sabbir}, \citenamefont {Smith}, \citenamefont {Wright}, \citenamefont {Rakher}, \citenamefont {Ladd},\ and\ \citenamefont {Borselli}}]{Weinstein2023}%
  \BibitemOpen
  \bibfield  {author} {\bibinfo {author} {\bibfnamefont {A.~J.}\ \bibnamefont {Weinstein}}, \bibinfo {author} {\bibfnamefont {M.~D.}\ \bibnamefont {Reed}}, \bibinfo {author} {\bibfnamefont {A.~M.}\ \bibnamefont {Jones}}, \bibinfo {author} {\bibfnamefont {R.~W.}\ \bibnamefont {Andrews}}, \bibinfo {author} {\bibfnamefont {D.}~\bibnamefont {Barnes}}, \bibinfo {author} {\bibfnamefont {J.~Z.}\ \bibnamefont {Blumoff}}, \bibinfo {author} {\bibfnamefont {L.~E.}\ \bibnamefont {Euliss}}, \bibinfo {author} {\bibfnamefont {K.}~\bibnamefont {Eng}}, \bibinfo {author} {\bibfnamefont {B.~H.}\ \bibnamefont {Fong}}, \bibinfo {author} {\bibfnamefont {S.~D.}\ \bibnamefont {Ha}}, \bibinfo {author} {\bibfnamefont {D.~R.}\ \bibnamefont {Hulbert}}, \bibinfo {author} {\bibfnamefont {C.~A.~C.}\ \bibnamefont {Jackson}}, \bibinfo {author} {\bibfnamefont {M.}~\bibnamefont {Jura}}, \bibinfo {author} {\bibfnamefont {T.~E.}\ \bibnamefont {Keating}}, \bibinfo {author} {\bibfnamefont {J.}~\bibnamefont {Kerckhoff}}, \bibinfo {author} {\bibfnamefont {A.~A.}\ \bibnamefont {Kiselev}}, \bibinfo {author} {\bibfnamefont {J.}~\bibnamefont {Matten}}, \bibinfo {author} {\bibfnamefont {G.}~\bibnamefont {Sabbir}}, \bibinfo {author} {\bibfnamefont {A.}~\bibnamefont {Smith}}, \bibinfo {author} {\bibfnamefont {J.}~\bibnamefont {Wright}}, \bibinfo {author} {\bibfnamefont {M.~T.}\ \bibnamefont {Rakher}}, \bibinfo {author} {\bibfnamefont {T.~D.}\ \bibnamefont {Ladd}},\ and\ \bibinfo {author} {\bibfnamefont {M.~G.}\ \bibnamefont {Borselli}},\ }\bibfield  {title} {\bibinfo {title} {Universal logic with encoded spin qubits in silicon},\ }\href {https://doi.org/10.1038/s41586-023-05777-3} {\bibfield  {journal} {\bibinfo  {journal} {Nature}\ }\textbf {\bibinfo {volume} {615}},\ \bibinfo {pages} {817} (\bibinfo {year} {2023})}\BibitemShut {NoStop}%
\bibitem [{\citenamefont {Hickman}\ \emph {et~al.}(2013)\citenamefont {Hickman}, \citenamefont {Wang}, \citenamefont {Kestner},\ and\ \citenamefont {Das~Sarma}}]{hickman2013}%
  \BibitemOpen
  \bibfield  {author} {\bibinfo {author} {\bibfnamefont {G.~T.}\ \bibnamefont {Hickman}}, \bibinfo {author} {\bibfnamefont {X.}~\bibnamefont {Wang}}, \bibinfo {author} {\bibfnamefont {J.~P.}\ \bibnamefont {Kestner}},\ and\ \bibinfo {author} {\bibfnamefont {S.}~\bibnamefont {Das~Sarma}},\ }\bibfield  {title} {\bibinfo {title} {Dynamically corrected gates for an exchange-only qubit},\ }\href {https://doi.org/10.1103/PhysRevB.88.161303} {\bibfield  {journal} {\bibinfo  {journal} {Phys. Rev. B}\ }\textbf {\bibinfo {volume} {88}},\ \bibinfo {pages} {161303} (\bibinfo {year} {2013})}\BibitemShut {NoStop}%
\bibitem [{\citenamefont {Shim}\ and\ \citenamefont {Tahan}(2016)}]{Shim2016}%
  \BibitemOpen
  \bibfield  {author} {\bibinfo {author} {\bibfnamefont {Y.-P.}\ \bibnamefont {Shim}}\ and\ \bibinfo {author} {\bibfnamefont {C.}~\bibnamefont {Tahan}},\ }\bibfield  {title} {\bibinfo {title} {Charge-noise-insensitive gate operations for always-on, exchange-only qubits},\ }\href {https://doi.org/10.1103/PhysRevB.93.121410} {\bibfield  {journal} {\bibinfo  {journal} {Phys. Rev. B}\ }\textbf {\bibinfo {volume} {93}},\ \bibinfo {pages} {121410} (\bibinfo {year} {2016})}\BibitemShut {NoStop}%
\bibitem [{\citenamefont {Shim}\ \emph {et~al.}(2025)\citenamefont {Shim}, \citenamefont {Takyi}, \citenamefont {Fei}, \citenamefont {Oh}, \citenamefont {Hu},\ and\ \citenamefont {Friesen}}]{Shim2013SinglequbitGI}%
  \BibitemOpen
  \bibfield  {author} {\bibinfo {author} {\bibfnamefont {Y.-P.}\ \bibnamefont {Shim}}, \bibinfo {author} {\bibfnamefont {E.}~\bibnamefont {Takyi}}, \bibinfo {author} {\bibfnamefont {J.}~\bibnamefont {Fei}}, \bibinfo {author} {\bibfnamefont {S.}~\bibnamefont {Oh}}, \bibinfo {author} {\bibfnamefont {X.}~\bibnamefont {Hu}},\ and\ \bibinfo {author} {\bibfnamefont {M.}~\bibnamefont {Friesen}},\ }\bibfield  {title} {\bibinfo {title} {Implementation of single-qubit gates with two rotations around axes in a plane},\ }\href {https://doi.org/10.1063/5.0267247} {\bibfield  {journal} {\bibinfo  {journal} {APL Quantum}\ }\textbf {\bibinfo {volume} {2}},\ \bibinfo {pages} {026122} (\bibinfo {year} {2025})}\BibitemShut {NoStop}%
\bibitem [{\citenamefont {Weinstein}\ and\ \citenamefont {Hellberg}(2005)}]{Weinstein2005}%
  \BibitemOpen
  \bibfield  {author} {\bibinfo {author} {\bibfnamefont {Y.~S.}\ \bibnamefont {Weinstein}}\ and\ \bibinfo {author} {\bibfnamefont {C.~S.}\ \bibnamefont {Hellberg}},\ }\bibfield  {title} {\bibinfo {title} {Energetic suppression of decoherence in exchange-only quantum computation},\ }\href {https://doi.org/10.1103/PhysRevA.72.022319} {\bibfield  {journal} {\bibinfo  {journal} {Phys. Rev. A}\ }\textbf {\bibinfo {volume} {72}},\ \bibinfo {pages} {022319} (\bibinfo {year} {2005})}\BibitemShut {NoStop}%
\bibitem [{\citenamefont {Doherty}\ and\ \citenamefont {Wardrop}(2013)}]{doherty2013_2qubit_rx_gates}%
  \BibitemOpen
  \bibfield  {author} {\bibinfo {author} {\bibfnamefont {A.~C.}\ \bibnamefont {Doherty}}\ and\ \bibinfo {author} {\bibfnamefont {M.~P.}\ \bibnamefont {Wardrop}},\ }\bibfield  {title} {\bibinfo {title} {Two-qubit gates for resonant exchange qubits},\ }\href {https://doi.org/10.1103/PhysRevLett.111.050503} {\bibfield  {journal} {\bibinfo  {journal} {Phys. Rev. Lett.}\ }\textbf {\bibinfo {volume} {111}},\ \bibinfo {pages} {050503} (\bibinfo {year} {2013})}\BibitemShut {NoStop}%
\bibitem [{\citenamefont {Hung}\ \emph {et~al.}(2014)\citenamefont {Hung}, \citenamefont {Fei}, \citenamefont {Friesen},\ and\ \citenamefont {Hu}}]{hung2014}%
  \BibitemOpen
  \bibfield  {author} {\bibinfo {author} {\bibfnamefont {J.-T.}\ \bibnamefont {Hung}}, \bibinfo {author} {\bibfnamefont {J.}~\bibnamefont {Fei}}, \bibinfo {author} {\bibfnamefont {M.}~\bibnamefont {Friesen}},\ and\ \bibinfo {author} {\bibfnamefont {X.}~\bibnamefont {Hu}},\ }\bibfield  {title} {\bibinfo {title} {Decoherence of an exchange qubit by hyperfine interaction},\ }\href {https://doi.org/10.1103/PhysRevB.90.045308} {\bibfield  {journal} {\bibinfo  {journal} {Phys. Rev. B}\ }\textbf {\bibinfo {volume} {90}},\ \bibinfo {pages} {045308} (\bibinfo {year} {2014})}\BibitemShut {NoStop}%
\bibitem [{\citenamefont {Acuna}\ \emph {et~al.}(2024)\citenamefont {Acuna}, \citenamefont {Broz}, \citenamefont {Shyamsundar}, \citenamefont {Mei}, \citenamefont {Feeney}, \citenamefont {Smetanka}, \citenamefont {Davis}, \citenamefont {Lee}, \citenamefont {Choi}, \citenamefont {Boyd}, \citenamefont {Suh}, \citenamefont {Ha}, \citenamefont {Jennings}, \citenamefont {Pan}, \citenamefont {Sanchez}, \citenamefont {Reed},\ and\ \citenamefont {Petta}}]{acuna2024}%
  \BibitemOpen
  \bibfield  {author} {\bibinfo {author} {\bibfnamefont {E.}~\bibnamefont {Acuna}}, \bibinfo {author} {\bibfnamefont {J.~D.}\ \bibnamefont {Broz}}, \bibinfo {author} {\bibfnamefont {K.}~\bibnamefont {Shyamsundar}}, \bibinfo {author} {\bibfnamefont {A.~B.}\ \bibnamefont {Mei}}, \bibinfo {author} {\bibfnamefont {C.~P.}\ \bibnamefont {Feeney}}, \bibinfo {author} {\bibfnamefont {V.}~\bibnamefont {Smetanka}}, \bibinfo {author} {\bibfnamefont {T.}~\bibnamefont {Davis}}, \bibinfo {author} {\bibfnamefont {K.}~\bibnamefont {Lee}}, \bibinfo {author} {\bibfnamefont {M.~D.}\ \bibnamefont {Choi}}, \bibinfo {author} {\bibfnamefont {B.}~\bibnamefont {Boyd}}, \bibinfo {author} {\bibfnamefont {J.}~\bibnamefont {Suh}}, \bibinfo {author} {\bibfnamefont {W.}~\bibnamefont {Ha}}, \bibinfo {author} {\bibfnamefont {C.}~\bibnamefont {Jennings}}, \bibinfo {author} {\bibfnamefont {A.~S.}\ \bibnamefont {Pan}}, \bibinfo {author} {\bibfnamefont {D.~S.}\ \bibnamefont {Sanchez}}, \bibinfo {author} {\bibfnamefont {M.~D.}\ \bibnamefont {Reed}},\ and\ \bibinfo {author} {\bibfnamefont {J.~R.}\ \bibnamefont {Petta}},\ }\bibfield  {title} {\bibinfo {title} {Coherent control of a triangular exchange-only spin qubit},\ }\href {https://doi.org/10.1103/PhysRevApplied.22.044057} {\bibfield  {journal} {\bibinfo  {journal} {Phys. Rev. Appl.}\ }\textbf {\bibinfo {volume} {22}},\ \bibinfo {pages} {044057} (\bibinfo {year} {2024})}\BibitemShut {NoStop}%
\bibitem [{\citenamefont {Russ}\ \emph {et~al.}(2016)\citenamefont {Russ}, \citenamefont {Ginzel},\ and\ \citenamefont {Burkard}}]{Russ2016}%
  \BibitemOpen
  \bibfield  {author} {\bibinfo {author} {\bibfnamefont {M.}~\bibnamefont {Russ}}, \bibinfo {author} {\bibfnamefont {F.}~\bibnamefont {Ginzel}},\ and\ \bibinfo {author} {\bibfnamefont {G.}~\bibnamefont {Burkard}},\ }\bibfield  {title} {\bibinfo {title} {Coupling of three-spin qubits to their electric environment},\ }\href {https://doi.org/10.1103/PhysRevB.94.165411} {\bibfield  {journal} {\bibinfo  {journal} {Phys. Rev. B}\ }\textbf {\bibinfo {volume} {94}},\ \bibinfo {pages} {165411} (\bibinfo {year} {2016})}\BibitemShut {NoStop}%
\bibitem [{\citenamefont {Hensgens}\ \emph {et~al.}(2017)\citenamefont {Hensgens}, \citenamefont {Fujita}, \citenamefont {Janssen}, \citenamefont {Li}, \citenamefont {Van~Diepen}, \citenamefont {Reichl}, \citenamefont {Wegscheider}, \citenamefont {Das~Sarma},\ and\ \citenamefont {Vandersypen}}]{Hensgens2017fermi-hubbard}%
  \BibitemOpen
  \bibfield  {author} {\bibinfo {author} {\bibfnamefont {T.}~\bibnamefont {Hensgens}}, \bibinfo {author} {\bibfnamefont {T.}~\bibnamefont {Fujita}}, \bibinfo {author} {\bibfnamefont {L.}~\bibnamefont {Janssen}}, \bibinfo {author} {\bibfnamefont {X.}~\bibnamefont {Li}}, \bibinfo {author} {\bibfnamefont {C.~J.}\ \bibnamefont {Van~Diepen}}, \bibinfo {author} {\bibfnamefont {C.}~\bibnamefont {Reichl}}, \bibinfo {author} {\bibfnamefont {W.}~\bibnamefont {Wegscheider}}, \bibinfo {author} {\bibfnamefont {S.}~\bibnamefont {Das~Sarma}},\ and\ \bibinfo {author} {\bibfnamefont {L.~M.~K.}\ \bibnamefont {Vandersypen}},\ }\bibfield  {title} {\bibinfo {title} {{Quantum simulation of a Fermi--Hubbard model using a semiconductor quantum dot array}},\ }\href {https://doi.org/10.1038/nature23022} {\bibfield  {journal} {\bibinfo  {journal} {Nature}\ }\textbf {\bibinfo {volume} {548}},\ \bibinfo {pages} {70} (\bibinfo {year} {2017})}\BibitemShut {NoStop}%
\bibitem [{\citenamefont {Mills}\ \emph {et~al.}(2019)\citenamefont {Mills}, \citenamefont {Feldman}, \citenamefont {Monical}, \citenamefont {Lewis}, \citenamefont {Larson}, \citenamefont {Mounce},\ and\ \citenamefont {Petta}}]{mills2019}%
  \BibitemOpen
  \bibfield  {author} {\bibinfo {author} {\bibfnamefont {A.~R.}\ \bibnamefont {Mills}}, \bibinfo {author} {\bibfnamefont {M.~M.}\ \bibnamefont {Feldman}}, \bibinfo {author} {\bibfnamefont {C.}~\bibnamefont {Monical}}, \bibinfo {author} {\bibfnamefont {P.~J.}\ \bibnamefont {Lewis}}, \bibinfo {author} {\bibfnamefont {K.~W.}\ \bibnamefont {Larson}}, \bibinfo {author} {\bibfnamefont {A.~M.}\ \bibnamefont {Mounce}},\ and\ \bibinfo {author} {\bibfnamefont {J.~R.}\ \bibnamefont {Petta}},\ }\bibfield  {title} {\bibinfo {title} {Computer-automated tuning procedures for semiconductor quantum dot arrays},\ }\href {https://doi.org/10.1063/1.5121444} {\bibfield  {journal} {\bibinfo  {journal} {Appl. Phys. Lett.}\ }\textbf {\bibinfo {volume} {115}},\ \bibinfo {pages} {113501} (\bibinfo {year} {2019})}\BibitemShut {NoStop}%
\bibitem [{\citenamefont {Hsiao}\ \emph {et~al.}(2020)\citenamefont {Hsiao}, \citenamefont {van Diepen}, \citenamefont {Mukhopadhyay}, \citenamefont {Reichl}, \citenamefont {Wegscheider},\ and\ \citenamefont {Vandersypen}}]{hsiao2020}%
  \BibitemOpen
  \bibfield  {author} {\bibinfo {author} {\bibfnamefont {T.-K.}\ \bibnamefont {Hsiao}}, \bibinfo {author} {\bibfnamefont {C.}~\bibnamefont {van Diepen}}, \bibinfo {author} {\bibfnamefont {U.}~\bibnamefont {Mukhopadhyay}}, \bibinfo {author} {\bibfnamefont {C.}~\bibnamefont {Reichl}}, \bibinfo {author} {\bibfnamefont {W.}~\bibnamefont {Wegscheider}},\ and\ \bibinfo {author} {\bibfnamefont {L.}~\bibnamefont {Vandersypen}},\ }\bibfield  {title} {\bibinfo {title} {Efficient orthogonal control of tunnel couplings in a quantum dot array},\ }\href {https://doi.org/10.1103/PhysRevApplied.13.054018} {\bibfield  {journal} {\bibinfo  {journal} {Phys. Rev. Appl.}\ }\textbf {\bibinfo {volume} {13}},\ \bibinfo {pages} {054018} (\bibinfo {year} {2020})}\BibitemShut {NoStop}%
\bibitem [{\citenamefont {Blumoff}\ \emph {et~al.}(2022)\citenamefont {Blumoff}, \citenamefont {Pan}, \citenamefont {Keating}, \citenamefont {Andrews}, \citenamefont {Barnes}, \citenamefont {Brecht}, \citenamefont {Croke}, \citenamefont {Euliss}, \citenamefont {Fast}, \citenamefont {Jackson}, \citenamefont {Jones}, \citenamefont {Kerckhoff}, \citenamefont {Lanza}, \citenamefont {Raach}, \citenamefont {Thomas}, \citenamefont {Velunta}, \citenamefont {Weinstein}, \citenamefont {Ladd}, \citenamefont {Eng}, \citenamefont {Borselli}, \citenamefont {Hunter},\ and\ \citenamefont {Rakher}}]{blumoff2022}%
  \BibitemOpen
  \bibfield  {author} {\bibinfo {author} {\bibfnamefont {J.~Z.}\ \bibnamefont {Blumoff}}, \bibinfo {author} {\bibfnamefont {A.~S.}\ \bibnamefont {Pan}}, \bibinfo {author} {\bibfnamefont {T.~E.}\ \bibnamefont {Keating}}, \bibinfo {author} {\bibfnamefont {R.~W.}\ \bibnamefont {Andrews}}, \bibinfo {author} {\bibfnamefont {D.~W.}\ \bibnamefont {Barnes}}, \bibinfo {author} {\bibfnamefont {T.~L.}\ \bibnamefont {Brecht}}, \bibinfo {author} {\bibfnamefont {E.~T.}\ \bibnamefont {Croke}}, \bibinfo {author} {\bibfnamefont {L.~E.}\ \bibnamefont {Euliss}}, \bibinfo {author} {\bibfnamefont {J.~A.}\ \bibnamefont {Fast}}, \bibinfo {author} {\bibfnamefont {C.~A.}\ \bibnamefont {Jackson}}, \bibinfo {author} {\bibfnamefont {A.~M.}\ \bibnamefont {Jones}}, \bibinfo {author} {\bibfnamefont {J.}~\bibnamefont {Kerckhoff}}, \bibinfo {author} {\bibfnamefont {R.~K.}\ \bibnamefont {Lanza}}, \bibinfo {author} {\bibfnamefont {K.}~\bibnamefont {Raach}}, \bibinfo {author} {\bibfnamefont {B.~J.}\ \bibnamefont {Thomas}}, \bibinfo {author} {\bibfnamefont {R.}~\bibnamefont {Velunta}}, \bibinfo {author} {\bibfnamefont {A.~J.}\ \bibnamefont {Weinstein}}, \bibinfo {author} {\bibfnamefont {T.~D.}\ \bibnamefont {Ladd}}, \bibinfo {author} {\bibfnamefont {K.}~\bibnamefont {Eng}}, \bibinfo {author} {\bibfnamefont {M.~G.}\ \bibnamefont {Borselli}}, \bibinfo {author} {\bibfnamefont {A.~T.}\ \bibnamefont {Hunter}},\ and\ \bibinfo {author} {\bibfnamefont {M.~T.}\ \bibnamefont {Rakher}},\ }\bibfield  {title} {\bibinfo {title} {Fast and high-fidelity state preparation and measurement in triple-quantum-dot spin qubits},\ }\href {https://doi.org/10.1103/PRXQuantum.3.010352} {\bibfield  {journal} {\bibinfo  {journal} {PRX Quantum}\ }\textbf {\bibinfo {volume} {3}},\ \bibinfo {pages} {010352} (\bibinfo {year} {2022})}\BibitemShut {NoStop}%
\bibitem [{\citenamefont {Lidar}\ \emph {et~al.}(1998)\citenamefont {Lidar}, \citenamefont {Chuang},\ and\ \citenamefont {Whaley}}]{Lidar1998}%
  \BibitemOpen
  \bibfield  {author} {\bibinfo {author} {\bibfnamefont {D.~A.}\ \bibnamefont {Lidar}}, \bibinfo {author} {\bibfnamefont {I.~L.}\ \bibnamefont {Chuang}},\ and\ \bibinfo {author} {\bibfnamefont {K.~B.}\ \bibnamefont {Whaley}},\ }\bibfield  {title} {\bibinfo {title} {Decoherence-free subspaces for quantum computation},\ }\href {https://doi.org/10.1103/PhysRevLett.81.2594} {\bibfield  {journal} {\bibinfo  {journal} {Phys. Rev. Lett.}\ }\textbf {\bibinfo {volume} {81}},\ \bibinfo {pages} {2594} (\bibinfo {year} {1998})}\BibitemShut {NoStop}%
\bibitem [{\citenamefont {Kempe}\ \emph {et~al.}(2001)\citenamefont {Kempe}, \citenamefont {Bacon}, \citenamefont {Lidar},\ and\ \citenamefont {Whaley}}]{kempe2001}%
  \BibitemOpen
  \bibfield  {author} {\bibinfo {author} {\bibfnamefont {J.}~\bibnamefont {Kempe}}, \bibinfo {author} {\bibfnamefont {D.}~\bibnamefont {Bacon}}, \bibinfo {author} {\bibfnamefont {D.~A.}\ \bibnamefont {Lidar}},\ and\ \bibinfo {author} {\bibfnamefont {K.~B.}\ \bibnamefont {Whaley}},\ }\bibfield  {title} {\bibinfo {title} {Theory of decoherence-free fault-tolerant universal quantum computation},\ }\href {https://doi.org/10.1103/PhysRevA.63.042307} {\bibfield  {journal} {\bibinfo  {journal} {Phys. Rev. A}\ }\textbf {\bibinfo {volume} {63}},\ \bibinfo {pages} {042307} (\bibinfo {year} {2001})}\BibitemShut {NoStop}%
\bibitem [{\citenamefont {Fong}\ and\ \citenamefont {Wandzura}(2011)}]{Fong2011}%
  \BibitemOpen
  \bibfield  {author} {\bibinfo {author} {\bibfnamefont {B.}~\bibnamefont {Fong}}\ and\ \bibinfo {author} {\bibfnamefont {S.}~\bibnamefont {Wandzura}},\ }\bibfield  {title} {\bibinfo {title} {Universal quantum computation and leakage reduction in the 3-qubit decoherence free subsystem},\ }\href {https://doi.org/10.26421/QIC11.11-12-9} {\bibfield  {journal} {\bibinfo  {journal} {Quantum Inf. Comput.}\ }\textbf {\bibinfo {volume} {11}},\ \bibinfo {pages} {1003} (\bibinfo {year} {2011})}\BibitemShut {NoStop}%
\bibitem [{\citenamefont {Ladd}(2012)}]{ladd2012}%
  \BibitemOpen
  \bibfield  {author} {\bibinfo {author} {\bibfnamefont {T.~D.}\ \bibnamefont {Ladd}},\ }\bibfield  {title} {\bibinfo {title} {Hyperfine-induced decay in triple quantum dots},\ }\href {https://doi.org/10.1103/PhysRevB.86.125408} {\bibfield  {journal} {\bibinfo  {journal} {Phys. Rev. B}\ }\textbf {\bibinfo {volume} {86}},\ \bibinfo {pages} {125408} (\bibinfo {year} {2012})}\BibitemShut {NoStop}%
\bibitem [{\citenamefont {Kerckhoff}\ \emph {et~al.}(2021)\citenamefont {Kerckhoff}, \citenamefont {Sun}, \citenamefont {Fong}, \citenamefont {Jones}, \citenamefont {Kiselev}, \citenamefont {Barnes}, \citenamefont {Noah}, \citenamefont {Acuna}, \citenamefont {Akmal}, \citenamefont {Ha}, \citenamefont {Wright}, \citenamefont {Thomas}, \citenamefont {Jackson}, \citenamefont {Edge}, \citenamefont {Eng}, \citenamefont {Ross},\ and\ \citenamefont {Ladd}}]{Kerckhoff2021}%
  \BibitemOpen
  \bibfield  {author} {\bibinfo {author} {\bibfnamefont {J.}~\bibnamefont {Kerckhoff}}, \bibinfo {author} {\bibfnamefont {B.}~\bibnamefont {Sun}}, \bibinfo {author} {\bibfnamefont {B.}~\bibnamefont {Fong}}, \bibinfo {author} {\bibfnamefont {C.}~\bibnamefont {Jones}}, \bibinfo {author} {\bibfnamefont {A.}~\bibnamefont {Kiselev}}, \bibinfo {author} {\bibfnamefont {D.}~\bibnamefont {Barnes}}, \bibinfo {author} {\bibfnamefont {R.}~\bibnamefont {Noah}}, \bibinfo {author} {\bibfnamefont {E.}~\bibnamefont {Acuna}}, \bibinfo {author} {\bibfnamefont {M.}~\bibnamefont {Akmal}}, \bibinfo {author} {\bibfnamefont {S.}~\bibnamefont {Ha}}, \bibinfo {author} {\bibfnamefont {J.}~\bibnamefont {Wright}}, \bibinfo {author} {\bibfnamefont {B.}~\bibnamefont {Thomas}}, \bibinfo {author} {\bibfnamefont {C.}~\bibnamefont {Jackson}}, \bibinfo {author} {\bibfnamefont {L.}~\bibnamefont {Edge}}, \bibinfo {author} {\bibfnamefont {K.}~\bibnamefont {Eng}}, \bibinfo {author} {\bibfnamefont {R.}~\bibnamefont {Ross}},\ and\ \bibinfo {author} {\bibfnamefont {T.}~\bibnamefont {Ladd}},\ }\bibfield  {title} {\bibinfo {title} {Magnetic gradient fluctuations from quadrupolar ${}^{73}\mathrm{Ge}$ in $\mathrm{Si}$/$\mathrm{Si}\mathrm{Ge}$ exchange-only qubits},\ }\href {https://doi.org/10.1103/PRXQuantum.2.010347} {\bibfield  {journal} {\bibinfo  {journal} {PRX Quantum}\ }\textbf {\bibinfo {volume} {2}},\ \bibinfo {pages} {010347} (\bibinfo {year} {2021})}\BibitemShut {NoStop}%
\bibitem [{\citenamefont {Heinz}\ \emph {et~al.}(2024)\citenamefont {Heinz}, \citenamefont {Borjans}, \citenamefont {Curry}, \citenamefont {Kotlyar}, \citenamefont {Luthi}, \citenamefont {Mądzik}, \citenamefont {Mohiyaddin}, \citenamefont {Bishop},\ and\ \citenamefont {Burkard}}]{heinz2024}%
  \BibitemOpen
  \bibfield  {author} {\bibinfo {author} {\bibfnamefont {I.}~\bibnamefont {Heinz}}, \bibinfo {author} {\bibfnamefont {F.}~\bibnamefont {Borjans}}, \bibinfo {author} {\bibfnamefont {M.}~\bibnamefont {Curry}}, \bibinfo {author} {\bibfnamefont {R.}~\bibnamefont {Kotlyar}}, \bibinfo {author} {\bibfnamefont {F.}~\bibnamefont {Luthi}}, \bibinfo {author} {\bibfnamefont {M.~T.}\ \bibnamefont {Mądzik}}, \bibinfo {author} {\bibfnamefont {F.~A.}\ \bibnamefont {Mohiyaddin}}, \bibinfo {author} {\bibfnamefont {N.}~\bibnamefont {Bishop}},\ and\ \bibinfo {author} {\bibfnamefont {G.}~\bibnamefont {Burkard}},\ }\bibfield  {title} {\bibinfo {title} {Fast quantum gates for exchange-only qubits using simultaneous exchange pulses,}\ }\href {https://doi.org/arXiv:2409.05843} {arXiv:2409.05843} (\bibinfo {year} {2024})\BibitemShut {NoStop}%
\bibitem [{\citenamefont {Qiao}\ \emph {et~al.}(2020)\citenamefont {Qiao}, \citenamefont {Kandel}, \citenamefont {Deng}, \citenamefont {Fallahi}, \citenamefont {Gardner}, \citenamefont {Manfra}, \citenamefont {Barnes},\ and\ \citenamefont {Nichol}}]{coherent_multispin_exchange}%
  \BibitemOpen
  \bibfield  {author} {\bibinfo {author} {\bibfnamefont {H.}~\bibnamefont {Qiao}}, \bibinfo {author} {\bibfnamefont {Y.~P.}\ \bibnamefont {Kandel}}, \bibinfo {author} {\bibfnamefont {K.}~\bibnamefont {Deng}}, \bibinfo {author} {\bibfnamefont {S.}~\bibnamefont {Fallahi}}, \bibinfo {author} {\bibfnamefont {G.~C.}\ \bibnamefont {Gardner}}, \bibinfo {author} {\bibfnamefont {M.~J.}\ \bibnamefont {Manfra}}, \bibinfo {author} {\bibfnamefont {E.}~\bibnamefont {Barnes}},\ and\ \bibinfo {author} {\bibfnamefont {J.~M.}\ \bibnamefont {Nichol}},\ }\bibfield  {title} {\bibinfo {title} {Coherent multispin exchange coupling in a quantum-dot spin chain},\ }\href {https://doi.org/10.1103/PhysRevX.10.031006} {\bibfield  {journal} {\bibinfo  {journal} {Phys. Rev. X}\ }\textbf {\bibinfo {volume} {10}},\ \bibinfo {pages} {031006} (\bibinfo {year} {2020})}\BibitemShut {NoStop}%
\bibitem [{\citenamefont {Fei}\ \emph {et~al.}(2015)\citenamefont {Fei}, \citenamefont {Hung}, \citenamefont {Koh}, \citenamefont {Shim}, \citenamefont {Coppersmith}, \citenamefont {Hu},\ and\ \citenamefont {Friesen}}]{Friesen2015}%
  \BibitemOpen
  \bibfield  {author} {\bibinfo {author} {\bibfnamefont {J.}~\bibnamefont {Fei}}, \bibinfo {author} {\bibfnamefont {J.-T.}\ \bibnamefont {Hung}}, \bibinfo {author} {\bibfnamefont {T.~S.}\ \bibnamefont {Koh}}, \bibinfo {author} {\bibfnamefont {Y.-P.}\ \bibnamefont {Shim}}, \bibinfo {author} {\bibfnamefont {S.~N.}\ \bibnamefont {Coppersmith}}, \bibinfo {author} {\bibfnamefont {X.}~\bibnamefont {Hu}},\ and\ \bibinfo {author} {\bibfnamefont {M.}~\bibnamefont {Friesen}},\ }\bibfield  {title} {\bibinfo {title} {Characterizing gate operations near the sweet spot of an exchange-only qubit},\ }\href {https://doi.org/10.1103/PhysRevB.91.205434} {\bibfield  {journal} {\bibinfo  {journal} {Phys. Rev. B}\ }\textbf {\bibinfo {volume} {91}},\ \bibinfo {pages} {205434} (\bibinfo {year} {2015})}\BibitemShut {NoStop}%
\bibitem [{\citenamefont {Malinowski}\ \emph {et~al.}(2017)\citenamefont {Malinowski}, \citenamefont {Martins}, \citenamefont {Nissen}, \citenamefont {Fallahi}, \citenamefont {Gardner}, \citenamefont {Manfra}, \citenamefont {Marcus},\ and\ \citenamefont {Kuemmeth}}]{Malinowski2017}%
  \BibitemOpen
  \bibfield  {author} {\bibinfo {author} {\bibfnamefont {F.~K.}\ \bibnamefont {Malinowski}}, \bibinfo {author} {\bibfnamefont {F.}~\bibnamefont {Martins}}, \bibinfo {author} {\bibfnamefont {P.~D.}\ \bibnamefont {Nissen}}, \bibinfo {author} {\bibfnamefont {S.}~\bibnamefont {Fallahi}}, \bibinfo {author} {\bibfnamefont {G.~C.}\ \bibnamefont {Gardner}}, \bibinfo {author} {\bibfnamefont {M.~J.}\ \bibnamefont {Manfra}}, \bibinfo {author} {\bibfnamefont {C.~M.}\ \bibnamefont {Marcus}},\ and\ \bibinfo {author} {\bibfnamefont {F.}~\bibnamefont {Kuemmeth}},\ }\bibfield  {title} {\bibinfo {title} {Symmetric operation of the resonant exchange qubit},\ }\href {https://doi.org/10.1103/PhysRevB.96.045443} {\bibfield  {journal} {\bibinfo  {journal} {Phys. Rev. B}\ }\textbf {\bibinfo {volume} {96}},\ \bibinfo {pages} {045443} (\bibinfo {year} {2017})}\BibitemShut {NoStop}%
\bibitem [{\citenamefont {Kimmel}\ \emph {et~al.}(2015)\citenamefont {Kimmel}, \citenamefont {Low},\ and\ \citenamefont {Yoder}}]{kimmel2015rpe}%
  \BibitemOpen
  \bibfield  {author} {\bibinfo {author} {\bibfnamefont {S.}~\bibnamefont {Kimmel}}, \bibinfo {author} {\bibfnamefont {G.~H.}\ \bibnamefont {Low}},\ and\ \bibinfo {author} {\bibfnamefont {T.~J.}\ \bibnamefont {Yoder}},\ }\bibfield  {title} {\bibinfo {title} {Robust calibration of a universal single-qubit gate set via robust phase estimation},\ }\href {https://doi.org/10.1103/PhysRevA.92.062315} {\bibfield  {journal} {\bibinfo  {journal} {Phys. Rev. A}\ }\textbf {\bibinfo {volume} {92}},\ \bibinfo {pages} {062315} (\bibinfo {year} {2015})}\BibitemShut {NoStop}%
\bibitem [{\citenamefont {Nielsen}\ \emph {et~al.}(2021)\citenamefont {Nielsen}, \citenamefont {Gamble}, \citenamefont {Rudinger}, \citenamefont {Scholten}, \citenamefont {Young},\ and\ \citenamefont {Blume-Kohout}}]{Nielsen2021gatesettomography}%
  \BibitemOpen
  \bibfield  {author} {\bibinfo {author} {\bibfnamefont {E.}~\bibnamefont {Nielsen}}, \bibinfo {author} {\bibfnamefont {J.~K.}\ \bibnamefont {Gamble}}, \bibinfo {author} {\bibfnamefont {K.}~\bibnamefont {Rudinger}}, \bibinfo {author} {\bibfnamefont {T.}~\bibnamefont {Scholten}}, \bibinfo {author} {\bibfnamefont {K.}~\bibnamefont {Young}},\ and\ \bibinfo {author} {\bibfnamefont {R.}~\bibnamefont {Blume-Kohout}},\ }\bibfield  {title} {\bibinfo {title} {Gate {S}et {T}omography},\ }\href {https://doi.org/10.22331/q-2021-10-05-557} {\bibfield  {journal} {\bibinfo  {journal} {{Quantum}}\ }\textbf {\bibinfo {volume} {5}},\ \bibinfo {pages} {557} (\bibinfo {year} {2021})}\BibitemShut {NoStop}%
\bibitem [{\citenamefont {Magesan}\ \emph {et~al.}(2012)\citenamefont {Magesan}, \citenamefont {Gambetta}, \citenamefont {Johnson}, \citenamefont {Ryan}, \citenamefont {Chow}, \citenamefont {Merkel}, \citenamefont {da~Silva}, \citenamefont {Keefe}, \citenamefont {Rothwell}, \citenamefont {Ohki}, \citenamefont {Ketchen},\ and\ \citenamefont {Steffen}}]{Magesan2012}%
  \BibitemOpen
  \bibfield  {author} {\bibinfo {author} {\bibfnamefont {E.}~\bibnamefont {Magesan}}, \bibinfo {author} {\bibfnamefont {J.~M.}\ \bibnamefont {Gambetta}}, \bibinfo {author} {\bibfnamefont {B.~R.}\ \bibnamefont {Johnson}}, \bibinfo {author} {\bibfnamefont {C.~A.}\ \bibnamefont {Ryan}}, \bibinfo {author} {\bibfnamefont {J.~M.}\ \bibnamefont {Chow}}, \bibinfo {author} {\bibfnamefont {S.~T.}\ \bibnamefont {Merkel}}, \bibinfo {author} {\bibfnamefont {M.~P.}\ \bibnamefont {da~Silva}}, \bibinfo {author} {\bibfnamefont {G.~A.}\ \bibnamefont {Keefe}}, \bibinfo {author} {\bibfnamefont {M.~B.}\ \bibnamefont {Rothwell}}, \bibinfo {author} {\bibfnamefont {T.~A.}\ \bibnamefont {Ohki}}, \bibinfo {author} {\bibfnamefont {M.~B.}\ \bibnamefont {Ketchen}},\ and\ \bibinfo {author} {\bibfnamefont {M.}~\bibnamefont {Steffen}},\ }\bibfield  {title} {\bibinfo {title} {Efficient measurement of quantum gate error by interleaved randomized benchmarking},\ }\href {https://doi.org/10.1103/PhysRevLett.109.080505} {\bibfield  {journal} {\bibinfo  {journal} {Phys. Rev. Lett.}\ }\textbf {\bibinfo {volume} {109}},\ \bibinfo {pages} {080505} (\bibinfo {year} {2012})}\BibitemShut {NoStop}%
\bibitem [{\citenamefont {Glaser}\ \emph {et~al.}(2015)\citenamefont {Glaser}, \citenamefont {Boscain}, \citenamefont {Calarco}, \citenamefont {Koch}, \citenamefont {K{\"o}ckenberger}, \citenamefont {Kosloff}, \citenamefont {Kuprov}, \citenamefont {Luy}, \citenamefont {Schirmer}, \citenamefont {Schulte-Herbr{\"u}ggen}, \citenamefont {Sugny},\ and\ \citenamefont {Wilhelm}}]{Glaser2015}%
  \BibitemOpen
  \bibfield  {author} {\bibinfo {author} {\bibfnamefont {S.~J.}\ \bibnamefont {Glaser}}, \bibinfo {author} {\bibfnamefont {U.}~\bibnamefont {Boscain}}, \bibinfo {author} {\bibfnamefont {T.}~\bibnamefont {Calarco}}, \bibinfo {author} {\bibfnamefont {C.~P.}\ \bibnamefont {Koch}}, \bibinfo {author} {\bibfnamefont {W.}~\bibnamefont {K{\"o}ckenberger}}, \bibinfo {author} {\bibfnamefont {R.}~\bibnamefont {Kosloff}}, \bibinfo {author} {\bibfnamefont {I.}~\bibnamefont {Kuprov}}, \bibinfo {author} {\bibfnamefont {B.}~\bibnamefont {Luy}}, \bibinfo {author} {\bibfnamefont {S.}~\bibnamefont {Schirmer}}, \bibinfo {author} {\bibfnamefont {T.}~\bibnamefont {Schulte-Herbr{\"u}ggen}}, \bibinfo {author} {\bibfnamefont {D.}~\bibnamefont {Sugny}},\ and\ \bibinfo {author} {\bibfnamefont {F.~K.}\ \bibnamefont {Wilhelm}},\ }\bibfield  {title} {\bibinfo {title} {{Training Schr{\"o}dinger's cat: quantum optimal control}},\ }\href {https://doi.org/10.1140/epjd/e2015-60464-1} {\bibfield  {journal} {\bibinfo  {journal} {Eur. Phys. J. D}\ }\textbf {\bibinfo {volume} {69}},\ \bibinfo {pages} {279} (\bibinfo {year} {2015})}\BibitemShut {NoStop}%
\bibitem [{\citenamefont {Gullans}\ and\ \citenamefont {Petta}(2019)}]{gullans2019protocol}%
  \BibitemOpen
  \bibfield  {author} {\bibinfo {author} {\bibfnamefont {M.}~\bibnamefont {Gullans}}\ and\ \bibinfo {author} {\bibfnamefont {J.}~\bibnamefont {Petta}},\ }\bibfield  {title} {\bibinfo {title} {{Protocol for a resonantly driven three-qubit Toffoli gate with silicon spin qubits}},\ }\href {https://doi.org/10.1103/PhysRevB.100.085419} {\bibfield  {journal} {\bibinfo  {journal} {Phys. Rev. B}\ }\textbf {\bibinfo {volume} {100}},\ \bibinfo {pages} {085419} (\bibinfo {year} {2019})}\BibitemShut {NoStop}%
\end{thebibliography}%


\begin{thebibliography}{10}%
\makeatletter
\providecommand \@ifxundefined [1]{%
 \@ifx{#1\undefined}
}%
\providecommand \@ifnum [1]{%
 \ifnum #1\expandafter \@firstoftwo
 \else \expandafter \@secondoftwo
 \fi
}%
\providecommand \@ifx [1]{%
 \ifx #1\expandafter \@firstoftwo
 \else \expandafter \@secondoftwo
 \fi
}%
\providecommand \natexlab [1]{#1}%
\providecommand \enquote  [1]{``#1''}%
\providecommand \bibnamefont  [1]{#1}%
\providecommand \bibfnamefont [1]{#1}%
\providecommand \citenamefont [1]{#1}%
\providecommand \href@noop [0]{\@secondoftwo}%
\providecommand \href [0]{\begingroup \@sanitize@url \@href}%
\providecommand \@href[1]{\@@startlink{#1}\@@href}%
\providecommand \@@href[1]{\endgroup#1\@@endlink}%
\providecommand \@sanitize@url [0]{\catcode `\\12\catcode `\$12\catcode `\&12\catcode `\#12\catcode `\^12\catcode `\_12\catcode `\%12\relax}%
\providecommand \@@startlink[1]{}%
\providecommand \@@endlink[0]{}%
\providecommand \url  [0]{\begingroup\@sanitize@url \@url }%
\providecommand \@url [1]{\endgroup\@href {#1}{\urlprefix }}%
\providecommand \urlprefix  [0]{URL }%
\providecommand \Eprint [0]{\href }%
\providecommand \doibase [0]{https://doi.org/}%
\providecommand \selectlanguage [0]{\@gobble}%
\providecommand \bibinfo  [0]{\@secondoftwo}%
\providecommand \bibfield  [0]{\@secondoftwo}%
\providecommand \translation [1]{[#1]}%
\providecommand \BibitemOpen [0]{}%
\providecommand \bibitemStop [0]{}%
\providecommand \bibitemNoStop [0]{.\EOS\space}%
\providecommand \EOS [0]{\spacefactor3000\relax}%
\providecommand \BibitemShut  [1]{\csname bibitem#1\endcsname}%
\let\auto@bib@innerbib\@empty
\bibitem [{Note1()}]{Note1}%
  \BibitemOpen
  \bibinfo {note} {For the gates calibrated in the main text, we used a 1-$J$ $\pi $ pulse about the $\protect \hat {z}$-axis to calibrate 2-$J$ rotations about the $\protect \hat {x}$-axis and used a 2-$J$ $\pi $ pulse about the $\protect \hat {x}$-axis to calibrate 2-$J$ rotations about the $-\protect \hat {z}$-axis.}\BibitemShut {Stop}%
\bibitem [{\citenamefont {Nielsen}\ \emph {et~al.}(2021)\citenamefont {Nielsen}, \citenamefont {Gamble}, \citenamefont {Rudinger}, \citenamefont {Scholten}, \citenamefont {Young},\ and\ \citenamefont {Blume-Kohout}}]{Nielsen2021gatesettomography}%
  \BibitemOpen
  \bibfield  {author} {\bibinfo {author} {\bibfnamefont {E.}~\bibnamefont {Nielsen}}, \bibinfo {author} {\bibfnamefont {J.~K.}\ \bibnamefont {Gamble}}, \bibinfo {author} {\bibfnamefont {K.}~\bibnamefont {Rudinger}}, \bibinfo {author} {\bibfnamefont {T.}~\bibnamefont {Scholten}}, \bibinfo {author} {\bibfnamefont {K.}~\bibnamefont {Young}},\ and\ \bibinfo {author} {\bibfnamefont {R.}~\bibnamefont {Blume-Kohout}},\ }\bibfield  {title} {\bibinfo {title} {Gate {S}et {T}omography},\ }\href {https://doi.org/10.22331/q-2021-10-05-557} {\bibfield  {journal} {\bibinfo  {journal} {{Quantum}}\ }\textbf {\bibinfo {volume} {5}},\ \bibinfo {pages} {557} (\bibinfo {year} {2021})}\BibitemShut {NoStop}%
\bibitem [{\citenamefont {Magesan}\ \emph {et~al.}(2011{\natexlab{a}})\citenamefont {Magesan}, \citenamefont {Gambetta},\ and\ \citenamefont {Emerson}}]{rb_paper}%
  \BibitemOpen
  \bibfield  {author} {\bibinfo {author} {\bibfnamefont {E.}~\bibnamefont {Magesan}}, \bibinfo {author} {\bibfnamefont {J.~M.}\ \bibnamefont {Gambetta}},\ and\ \bibinfo {author} {\bibfnamefont {J.}~\bibnamefont {Emerson}},\ }\bibfield  {title} {\bibinfo {title} {Scalable and robust randomized benchmarking of quantum processes},\ }\href {https://doi.org/10.1103/PhysRevLett.106.180504} {\bibfield  {journal} {\bibinfo  {journal} {Phys. Rev. Lett.}\ }\textbf {\bibinfo {volume} {106}},\ \bibinfo {pages} {180504} (\bibinfo {year} {2011}{\natexlab{a}})}\BibitemShut {NoStop}%
\bibitem [{\citenamefont {Magesan}\ \emph {et~al.}(2011{\natexlab{b}})\citenamefont {Magesan}, \citenamefont {Blume-Kohout},\ and\ \citenamefont {Emerson}}]{magesan2011}%
  \BibitemOpen
  \bibfield  {author} {\bibinfo {author} {\bibfnamefont {E.}~\bibnamefont {Magesan}}, \bibinfo {author} {\bibfnamefont {R.}~\bibnamefont {Blume-Kohout}},\ and\ \bibinfo {author} {\bibfnamefont {J.}~\bibnamefont {Emerson}},\ }\bibfield  {title} {\bibinfo {title} {Gate fidelity fluctuations and quantum process invariants},\ }\href {https://doi.org/10.1103/PhysRevA.84.012309} {\bibfield  {journal} {\bibinfo  {journal} {Phys. Rev. A}\ }\textbf {\bibinfo {volume} {84}},\ \bibinfo {pages} {012309} (\bibinfo {year} {2011}{\natexlab{b}})}\BibitemShut {NoStop}%
\bibitem [{Note2()}]{Note2}%
  \BibitemOpen
  \bibinfo {note} {Sequences corresponding to different $\protect \hat {C}_i$ are randomly ordered and typically we repeat each measurement for $\sim $10 shots to estimate the probability}\BibitemShut {NoStop}%
\bibitem [{Note3()}]{Note3}%
  \BibitemOpen
  \bibinfo {note} {We initialize the procedure using a coarse estimate of the peak location based on ``fingerpinch'' sweeps similar to those shown in Fig. 2 of the main text.}\BibitemShut {Stop}%
\bibitem [{\citenamefont {Kimmel}\ \emph {et~al.}(2015)\citenamefont {Kimmel}, \citenamefont {Low},\ and\ \citenamefont {Yoder}}]{kimmel2015rpe}%
  \BibitemOpen
  \bibfield  {author} {\bibinfo {author} {\bibfnamefont {S.}~\bibnamefont {Kimmel}}, \bibinfo {author} {\bibfnamefont {G.~H.}\ \bibnamefont {Low}},\ and\ \bibinfo {author} {\bibfnamefont {T.~J.}\ \bibnamefont {Yoder}},\ }\bibfield  {title} {\bibinfo {title} {Robust calibration of a universal single-qubit gate set via robust phase estimation},\ }\href {https://doi.org/10.1103/PhysRevA.92.062315} {\bibfield  {journal} {\bibinfo  {journal} {Phys. Rev. A}\ }\textbf {\bibinfo {volume} {92}},\ \bibinfo {pages} {062315} (\bibinfo {year} {2015})}\BibitemShut {NoStop}%
\bibitem [{\citenamefont {Wildberger}(2005)}]{wildberger2005divine}%
  \BibitemOpen
  \bibfield  {author} {\bibinfo {author} {\bibfnamefont {N.~J.}\ \bibnamefont {Wildberger}},\ }\href@noop {} {\emph {\bibinfo {title} {Divine Proportions: Rational Trigonometry to Universal Geometry}}}\ (\bibinfo  {publisher} {Wild Egg Books},\ \bibinfo {address} {Sydney, Australia},\ \bibinfo {year} {2005})\BibitemShut {NoStop}%
\bibitem [{\citenamefont {Andrews}\ \emph {et~al.}(2019)\citenamefont {Andrews}, \citenamefont {Jones}, \citenamefont {Reed}, \citenamefont {Jones}, \citenamefont {Ha}, \citenamefont {Jura}, \citenamefont {Kerckhoff}, \citenamefont {Levendorf}, \citenamefont {Meenehan}, \citenamefont {Merkel}, \citenamefont {Smith}, \citenamefont {Sun}, \citenamefont {Weinstein}, \citenamefont {Rakher}, \citenamefont {Ladd},\ and\ \citenamefont {Borselli}}]{Andrews2019}%
  \BibitemOpen
  \bibfield  {author} {\bibinfo {author} {\bibfnamefont {R.~W.}\ \bibnamefont {Andrews}}, \bibinfo {author} {\bibfnamefont {C.}~\bibnamefont {Jones}}, \bibinfo {author} {\bibfnamefont {M.~D.}\ \bibnamefont {Reed}}, \bibinfo {author} {\bibfnamefont {A.~M.}\ \bibnamefont {Jones}}, \bibinfo {author} {\bibfnamefont {S.~D.}\ \bibnamefont {Ha}}, \bibinfo {author} {\bibfnamefont {M.~P.}\ \bibnamefont {Jura}}, \bibinfo {author} {\bibfnamefont {J.}~\bibnamefont {Kerckhoff}}, \bibinfo {author} {\bibfnamefont {M.}~\bibnamefont {Levendorf}}, \bibinfo {author} {\bibfnamefont {S.}~\bibnamefont {Meenehan}}, \bibinfo {author} {\bibfnamefont {S.~T.}\ \bibnamefont {Merkel}}, \bibinfo {author} {\bibfnamefont {A.}~\bibnamefont {Smith}}, \bibinfo {author} {\bibfnamefont {B.}~\bibnamefont {Sun}}, \bibinfo {author} {\bibfnamefont {A.~J.}\ \bibnamefont {Weinstein}}, \bibinfo {author} {\bibfnamefont {M.~T.}\ \bibnamefont {Rakher}}, \bibinfo {author} {\bibfnamefont {T.~D.}\ \bibnamefont {Ladd}},\ and\ \bibinfo {author} {\bibfnamefont {M.~G.}\ \bibnamefont {Borselli}},\ }\bibfield  {title} {\bibinfo {title} {{Quantifying error and leakage in an encoded Si/SiGe triple-dot qubit}},\ }\href {https://doi.org/10.1038/s41565-019-0500-4} {\bibfield  {journal} {\bibinfo  {journal} {Nat. Nanotechnol.}\ }\textbf {\bibinfo {volume} {14}},\ \bibinfo {pages} {747} (\bibinfo {year} {2019})}\BibitemShut {NoStop}%
\bibitem [{\citenamefont {Sun}\ \emph {et~al.}(2024)\citenamefont {Sun}, \citenamefont {Brecht}, \citenamefont {Fong}, \citenamefont {Akmal}, \citenamefont {Blumoff}, \citenamefont {Cain}, \citenamefont {Carter}, \citenamefont {Finestone}, \citenamefont {Fireman}, \citenamefont {Ha}, \citenamefont {Hatke}, \citenamefont {Hickey}, \citenamefont {Jackson}, \citenamefont {Jenkins}, \citenamefont {Jones}, \citenamefont {Pan}, \citenamefont {Ward}, \citenamefont {Weinstein}, \citenamefont {Whiteley}, \citenamefont {Williams}, \citenamefont {Borselli}, \citenamefont {Rakher},\ and\ \citenamefont {Ladd}}]{sun2024}%
  \BibitemOpen
  \bibfield  {author} {\bibinfo {author} {\bibfnamefont {B.}~\bibnamefont {Sun}}, \bibinfo {author} {\bibfnamefont {T.}~\bibnamefont {Brecht}}, \bibinfo {author} {\bibfnamefont {B.~H.}\ \bibnamefont {Fong}}, \bibinfo {author} {\bibfnamefont {M.}~\bibnamefont {Akmal}}, \bibinfo {author} {\bibfnamefont {J.~Z.}\ \bibnamefont {Blumoff}}, \bibinfo {author} {\bibfnamefont {T.~A.}\ \bibnamefont {Cain}}, \bibinfo {author} {\bibfnamefont {F.~W.}\ \bibnamefont {Carter}}, \bibinfo {author} {\bibfnamefont {D.~H.}\ \bibnamefont {Finestone}}, \bibinfo {author} {\bibfnamefont {M.~N.}\ \bibnamefont {Fireman}}, \bibinfo {author} {\bibfnamefont {W.}~\bibnamefont {Ha}}, \bibinfo {author} {\bibfnamefont {A.~T.}\ \bibnamefont {Hatke}}, \bibinfo {author} {\bibfnamefont {R.~M.}\ \bibnamefont {Hickey}}, \bibinfo {author} {\bibfnamefont {C.~A.~C.}\ \bibnamefont {Jackson}}, \bibinfo {author} {\bibfnamefont {I.}~\bibnamefont {Jenkins}}, \bibinfo {author} {\bibfnamefont {A.~M.}\ \bibnamefont {Jones}}, \bibinfo {author} {\bibfnamefont {A.}~\bibnamefont {Pan}}, \bibinfo {author} {\bibfnamefont {D.~R.}\ \bibnamefont {Ward}}, \bibinfo {author} {\bibfnamefont {A.~J.}\ \bibnamefont {Weinstein}}, \bibinfo {author} {\bibfnamefont {S.~J.}\ \bibnamefont {Whiteley}}, \bibinfo {author} {\bibfnamefont {P.}~\bibnamefont {Williams}}, \bibinfo {author} {\bibfnamefont {M.~G.}\ \bibnamefont {Borselli}}, \bibinfo {author} {\bibfnamefont {M.~T.}\ \bibnamefont {Rakher}},\ and\ \bibinfo {author} {\bibfnamefont {T.~D.}\ \bibnamefont {Ladd}},\ }\bibfield  {title} {\bibinfo {title} {Full-permutation dynamical decoupling in triple-quantum-dot spin qubits},\ }\href {https://doi.org/10.1103/PRXQuantum.5.020356} {\bibfield  {journal} {\bibinfo  {journal} {PRX Quantum}\ }\textbf {\bibinfo {volume} {5}},\ \bibinfo {pages} {020356} (\bibinfo {year} {2024})}\BibitemShut {NoStop}%
\end{thebibliography}%

\section{Methods}

\subsection{Virtual gates}
Virtual gate voltages $\mathbf{\tilde{V}}$ are related to physical gate voltages $\mathbf{V}$ through a compensation matrix $C$, such that ${\mathbf{\tilde{V}}=C\mathbf{V}}$ \cite{Hensgens2017fermi-hubbard,mills2019,hsiao2020}. Ideally, the chemical potential $\epsilon_i$ is only affected by a voltage $V_{P_i}$ applied to the the plunger gate $P_i$. In reality, voltages applied to neighboring gates will also affect $\epsilon_i$. The virtual gates, as defined by $C$, are designed to compensate for this cross-coupling to first order. Explicitly, in this work we used:
\begin{equation}
	\begin{pmatrix}
	\tilde{V}_{P_{1}} \\ \tilde{V}_{P_{2}} \\ \tilde{V}_{P_{3}} \\ \tilde{V}_{X_{12}} \\ \tilde{V}_{X_{13}} \\ \tilde{V}_{X_{23}}
	\end{pmatrix} = 
	\begin{pmatrix}
	1 & 0.19 & 0.18 & 0.51 & 0.67 & 0.21 \\
	-0.19 & 1 & 0.20 & 0.38 & 0.36 & 0.49 \\
	0.06 & 0.20 & 1 & 0.16 & 0.98 & 0.53 \\
	0 & 0 & 0 & 1 & 0 & 0 \\
	0 & 0 & 0 & 0 & 1 & 0 \\
	0 & 0 & 0 & 0 & 0 & 1
	\end{pmatrix}
	\begin{pmatrix}
		V_{P_{1}} \\ V_{P_{2}} \\ V_{P_{3}} \\ V_{X_{12}} \\ V_{X_{13}} \\ V_{X_{23}}
	\end{pmatrix}.
\end{equation}

\noindent Here, the matrix elements are defined as $C_{i,j}=(\partial \epsilon_i / \partial V_{P_j}) / (\partial \epsilon_i / \partial V_{P_i}) $ with analogous definitions for exchange gates. We determine the values of the $C_{i,j}$ by tracking shifts of electron loading lines for each dot as a function of gate voltage. 

In the Fig.~\ref{fig:figure2} data, we also compensated for first-order cross-coupling between exchange gates using:

\begin{align}
	\begin{pmatrix}
		\tilde{V}'_{X_{1,2}} \\ \tilde{V}'_{X_{1,3}} \\ \tilde{V}'_{X_{2,3}}
	\end{pmatrix} &= 
	\begin{pmatrix}
		1 & d_{12,13} & d_{12,23} \\
		d_{13,12} & 1 & d_{13,23} \\
		d_{23,12} & d_{23,13} & 1
	\end{pmatrix}
	\begin{pmatrix}
		\tilde{V}_{X_{1,2}} \\ \tilde{V}_{X_{1,3}} \\ \tilde{V}_{X_{2,3}}
	\end{pmatrix} \\
	&= 
	\begin{pmatrix}
		1 & -0.08 & -0.08 \\
		-0.24 & 1 & -0.18 \\
		-0.15 & -0.19 & 1
	\end{pmatrix}
	\begin{pmatrix}
		\tilde{V}_{X_{1,2}} \\ \tilde{V}_{X_{1,3}} \\ \tilde{V}_{X_{2,3}}
	\end{pmatrix},
\end{align}

\noindent where $d_{ij,kl}=(\partial J_{i,j}/\partial \tilde{V}_{X_{k,l}})/ (\partial J_{i,j}/\partial \tilde{V}_{X_{i,j}})$. Here, the $d_{ij,kl}$ were determined by tracking the first exchange fringe in the fingerpinch plots as a function of neighboring exchange gate voltage near the center of the sweep. To simplify notation, we drop the superscripts in the labels of the virtual exchange gates in Fig.~\ref{fig:figure2}. We emphasize that exchange gate compensation was not used in any of the other data presented in this work (in particular, it was not applied for the calibration sweeps shown in Fig.~\ref{fig:figure4}).

\section{Data availability}
The data that supports the findings of this study are available from the corresponding authors upon reasonable request.

\section{Acknowledgments}
We thank John B. Carpenter for assisting with the preparation of the figures, Quantum Machines for access to the hardware used to perform the experiments (QDAC-II and OPX+), and Minh Nguyen for logistical support. The sample used in this experiment was made by the HRL device fabrication team. Research was supported by Army Research Office (ARO) grants W911NF-24-1-0020 and W911NF-22-C-0002. The views and conclusions contained in this document are those of the authors and should not be interpreted as representing the official policies, either expressed or implied, of the ARO or the U.S. Government. The U.S. Government is authorized to reproduce and distribute reprints for Government purposes notwithstanding any copyright notation herein.

\section{Author contributions}
J.D.B. and E.A. conceptualized the experiment. J.D.B. designed the experiments and collected the data. J.D.B. and J.C.H. analyzed the data and prepared the figures with input from J.R.P. J.R.P. designed the device. J.R.P. and E.A. supervised the project. All authors contributed to the writing of the manuscript.

\section{Competing interests}
J.R.P. has a significant financial interest in HRL Laboratories, LLC.

\end{document}



\title{Supplementary Information for: \\ Demonstration of an always-on exchange-only spin qubit}

\author{Joseph D. Broz}
\email[]{jdbroz@hrl.com}
\affiliation{HRL Laboratories, LLC, 3011 Malibu Canyon Road, Malibu, California 90265, USA}
\author{Jesse C. Hoke}
\affiliation{HRL Laboratories, LLC, 3011 Malibu Canyon Road, Malibu, California 90265, USA}
\author{Edwin Acuna}
\affiliation{HRL Laboratories, LLC, 3011 Malibu Canyon Road, Malibu, California 90265, USA}
\author{Jason R. Petta}
\email[]{jpetta1@hrl.com}
\affiliation{HRL Laboratories, LLC, 3011 Malibu Canyon Road, Malibu, California 90265, USA}
\affiliation{Department of Physics and Astronomy, University of California -- Los Angeles, Los Angeles, California 90095, USA}
\affiliation{Center for Quantum Science and Engineering, University of California -- Los Angeles, Los Angeles, California 90095, USA}

\maketitle

\section{Device tune-up}

\begin{figure*}[h]
	\renewcommand{\thefigure}{S\arabic{figure}}
	\centering
	\includegraphics[width=0.99\textwidth]{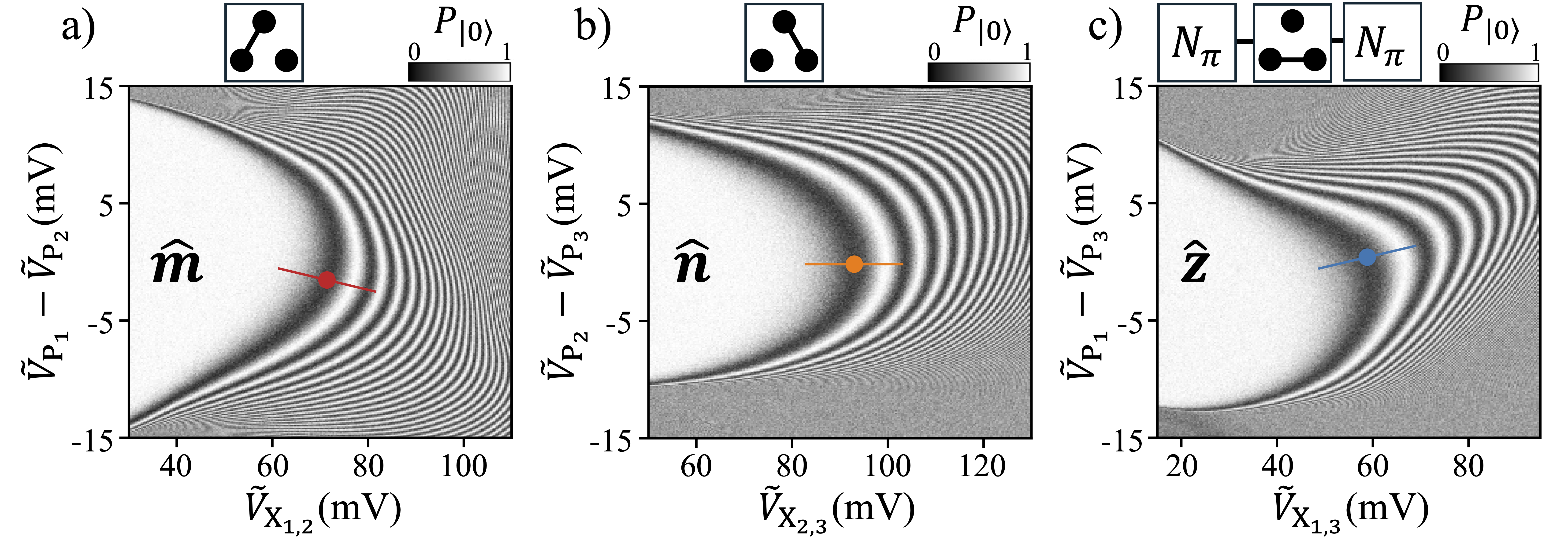}
	\caption{\textbf{1-$J$ exchange fingerprints}. (a) -- (c) 1-$J$ ``Fingerprint'' plots for the $\hat{m}$, $\hat{n}$, and $\hat{z}$ axes. Colored circles indicate the approximate locations of the 1-$J$ sweet spots where the exchange coupling is first-order insensitive to chemical potential fluctuations. The 1-$J$ exchange rotations used throughout this work were calibrated along the symmetric axes of operations (colored lines), along which this condition is maintained. Pre/post $\pi$-rotations about the $\hat{n}$-axis are applied in (c).
	} 
	\label{fig:fingerprint}
\end{figure*}

1-$J$ control with each of the exchange axes is illustrated in Fig.~\ref{fig:fingerprint}. We plot the probability to return to state 0, $P_{|0>}$, as a function of virtual plunger gate detunings (e.g. $\tilde{V}_{P_1}$~$-$~$\tilde{V}_{P_2}$) and virtual exchange gate voltages (e.g.~$\tilde{V}_{X_{1,2}}$). The $P_{|0>}$ oscillation frequency increases with $\tilde{V}_{X_{i,j}}$, as expected

\section{2-$J$ gate calibration}


The Hamiltonian for an exchange-only (EO) qubit can be written as:

\begin{equation} \label{eq;H_exc qubit operators}
    \hat{H} = -\frac{1}{2}[\sqrt{3}J_-\hat{\sigma}_x + (J_{1,3} - J_+)\hat{\sigma}_z],
\end{equation}

\noindent where $\hat{\sigma}_i$ are the standard Pauli operators, and $J_+=(J_{1,2} + J_{2,3})/2$ and $J_-=(J_{1,2} - J_{2,3})/2$. An exchange pulse of duration $\tau$ generates a rotation by an angle of $\theta$ about an axis $\mathbf{r}= (r_x, r_y, r_z) = (\textrm{cos}(\varphi),0,\textrm{sin}(\varphi))$ in the $xz$-plane of the Bloch sphere, as described by the unitary operator: 
\begin{equation}
    \hat{R}_{\varphi}(\theta)\approx \textrm{cos}(\theta/2) - i\textrm{sin}(\theta/2)\mathbf{r}\cdot\hat{\boldsymbol{\sigma}},\label{eq:qubit rotation}
\end{equation}
where $\hat{\boldsymbol{\sigma}} = (\hat{\sigma_x}, \hat{\sigma_y}, \hat{\sigma_z})$, $\theta=\int_0^\tau \Omega(t) dt$, and:
\begin{align}
    \Omega &= \sqrt{3J_-^2 + (J_{1,3} - J_+)^2}, \label{eq:Omega}\\
    \textrm{cos}(\varphi) &= \sqrt{3}J_-/\Omega, \label{eq:cos(phi)}\\
    \textrm{sin}(\varphi) &= (J_{1,3} - J_+) / \Omega \label{eq:sin(phi)}.
\end{align}
The values of $\theta$ and $\varphi$ are determined by the strength, $J_{i,j}$, of the pairwise exchange interactions activated during the pulse (Eqs. \ref{eq:Omega} -- \ref{eq:sin(phi)}), which, in turn, depend on the amplitude of the voltage pulses applied to the gate electrodes. For 1-$J$ exchange, $\varphi$ is restricted to the discrete set $\{0,2\pi/3,4\pi/3\}$, corresponding to the $\{\hat{z},\hat{n},\hat{m}\}$ axes shown in Fig.~1(b) of the main text. In contrast, for 2-$J$ exchange pulses, $\varphi$ can be tuned to an arbitrary value.

To calibrate a 2-$J$ rotation $\hat{R}_{\varphi^*}(\theta^*)$, corresponding to target angles $\varphi^*$ and $\theta^*$, we employ the composite sequence $\hat{U}(N)$, where:
\begin{align}
	\hat{U}(N) &= \hat{U}_{\textrm{ax}}^N\hat{U}_{\textrm{ang}}^N, \\
	\hat{U}_{\textrm{ax}} &= \hat{R}_{\varphi}^{2q}(\theta), \\
	\hat{U}_{\textrm{ang}} &= (\hat{R}_{\eta}(\chi)\hat{R}_{\varphi}^q(\theta))^2,
\end{align}
\noindent and the integer $q$ is chosen such that $q\theta^*=s\pi$ for some odd integer $s$. The operator $\hat{R}_{\eta}(\chi)$ is a pre-calibrated rotation with parameters $\eta\approx\varphi^*+\pi/2$ and $\chi\approx\pi$, and is assumed to be implemented using a single 1- or 2-$J$ exchange pulse \footnote{For the gates calibrated in the main text, we used a 1-$J$ $\pi$ pulse about the $\hat{z}$-axis to calibrate 2-$J$ rotations about the $\hat{x}$-axis and used a 2-$J$ $\pi$ pulse about the $\hat{x}$-axis to calibrate 2-$J$ rotations about the $-\hat{z}$-axis.}. The sequence $\hat{U}(N)$ is designed to reduce to the identity when $\hat{R}_{\varphi}(\theta)$ is perfectly calibrated, i.e., 
\begin{equation} \label{eq:optimal calibration condition}
\hat{U}(N;\varphi^*,\theta^*)=\hat{\mathbb{1}}.
\end{equation} 
\noindent Furthermore, $\hat{U}_{\textrm{ax}}$ is designed to amplify deviations from Eq.~\ref{eq:optimal calibration condition} due to errors $\epsilon_{\varphi}$ in the axis angle, $\varphi=\varphi^*+\epsilon_{\varphi}$, by a factor that scales with the repetition number $N$. While $\hat{U}_{\textrm{ang}}$ similarly amplifies errors $\epsilon_{\theta}$ in the rotation angle $\theta$. In the terminology of gate set tomography, $\hat{U}_{\textrm{ax}}$ and $\hat{U}_{\textrm{ang}}$ are called \textit{germs} and $N$ the \textit{germ power} \cite{Nielsen2021gatesettomography}. By twirling $\hat{U}$ over the set of 1-$J$ single-qubit Clifford gates $\mathcal{C}$, we can directly measure the fidelity, $F(\hat{U}, \hat{\mathbb{1}}$), of $\hat{U}$ relative to the identity \cite{rb_paper,magesan2011}:
\begin{equation}
	\label{eq:fidelity}
	F(\hat{U}, \hat{\mathbb{1}}) = \frac{1}{|\mathcal{C}|}\sum_{i=1}^{|\mathcal{C}|} |\langle 0 | \hat{C}_i^{\dagger} \hat{U} \hat{C}_i | 0 \rangle|^2.
\end{equation}
\noindent In practice, each term in the summation is obtained by initializing the qubit in $\ket{0}$, applying the sequence $\hat{C}_i^{\dagger} \hat{U} \hat{C}_i$, and then measuring the probability that the qubit remains in $\ket{0}$ \footnote{Sequences corresponding to different $\hat{C}_i$ are randomly ordered and typically we repeat each measurement for $\sim$10 shots to estimate the probability}. By measuring $F$ across a range of the exchange gate voltages used to implement $\hat{R}_{\varphi}(\theta)$ we generate two-dimensional plots similar to those shown in Fig.~4 of the main text. These data exhibit a series of peaks, but only the central peak corresponds to the optimal calibration point where $\varphi=\varphi^*$ and $\theta=\theta^*$. To track this peak, we perform successive sweeps with increasing values of $N=1,2,4,8,16,\ldots$, continuing until reductions in the signal-to-noise ratio prevents further scaling.  Practically, charge noise limits us from setting $N$ much larger, but we note that already with $N = 24$ we are able to estimate the peak bias values at a precision comparable to the limitations of our control hardware ($\sim 7$ $\mu$V). At each stage, we select the peak nearest to the value identified in the previous step \footnote{We initialize the procedure using a coarse estimate of the peak location based on ``fingerpinch'' sweeps similar to those shown in Fig. 2 of the main text.}. This iterative approach is similar to the approach used in robust phase estimation \cite{kimmel2015rpe}, and the entire protocol can be viewed as a two-dimensional generalization of that technique. While not strictly necessary, we find that twirling $\hat{U}$ enhances the contrast of the interference peaks and suppresses the effects of time-correlated noise.

To accurately determine the location of the central calibration peak using this protocol, we fit the measured data to the analytical expression (e.g. red contours in Fig. 4 of the main text):
\begin{align} \label{eq:F}
\begin{split}
F(\hat{U}, \hat{\mathbb{1}}) = 1 - \frac{2}{3}\bigg\{&\bigg[\cos(N\theta)
+ \bigg((r_xk_z - r_zk_x)^2 + k_y^2\bigg) \sin^2(N\theta/2)\bigg]S_{2N}[\sin^2(\Phi/2)] \\
&+ \frac{1}{2}(r_xk_x + r_zk_z)\sin(N\theta)U_{4N-1}[\cos(\Phi/2)]\sin(\Phi/2) + \sin^2(N\theta/2)
\bigg\}.
\end{split}
\end{align}
\noindent Here, $S_{M}$ are spread polynomials of order $M$ \cite{wildberger2005divine}, which can be expressed in terms of the Chebyshev polynomials of the first kind as $T_M$ as $S_{M}(x) = [1 - T_{M}(1 - 2x)]/2$, and $U_{M}$ are Chebyshev polynomials of the second kind. The parameters $\Phi$ and $\mathbf{k}=(k_x,k_y,k_z)$ characterize the net rotation resulting from the composition $\hat{R}_{\chi}(\eta)\hat{R}_{\varphi}(\theta)$:
\begin{equation}
	\hat{R}_{\chi}(\eta)\hat{R}_{\varphi}(\theta) = \cos(\Phi/2) - i\sin(\Phi/2)\mathbf{k}\cdot\hat{\boldsymbol{\sigma}},
\end{equation}
\noindent with explicit expressions:
\begin{align}
\cos(\Phi/2) &= \cos(\chi/2)\cos(\theta/2) - \sin(\chi/2)\sin(\theta/2)\cos(\varphi-\eta), \\
\sin^2(\Phi/2) &= \textrm{cos}(\chi/2)\textrm{sin}(\theta/2)+\textrm{sin}(\chi/2)\textrm{cos}(\theta/2)\textrm{cos}^2(\varphi-\eta) +\textrm{sin}^2(\chi/2)\textrm{sin}^2(\varphi-\eta), \\
k_x &= \frac{\cos(\varphi)\cos(\chi/2)\sin(\theta/2) + \sin(\chi/2)\cos(\theta/2)\cos(\eta)}{\sin(\Phi/2)}, \\
k_y &= -\frac{\sin(\theta/2)\sin(\chi/2)\sin(\varphi-\eta)}{\sin(\Phi/2)}, {\rm and} \\
k_z &= \frac{\sin(\chi/2)\cos(\theta/2)\sin(\eta) + \sin(\varphi)\cos(\chi/2)\sin(\theta/2)}{\sin(\Phi/2)}.
\end{align}

\begin{figure*}
	\renewcommand{\thefigure}{S\arabic{figure}}
\centering
\includegraphics[width=0.8\textwidth]{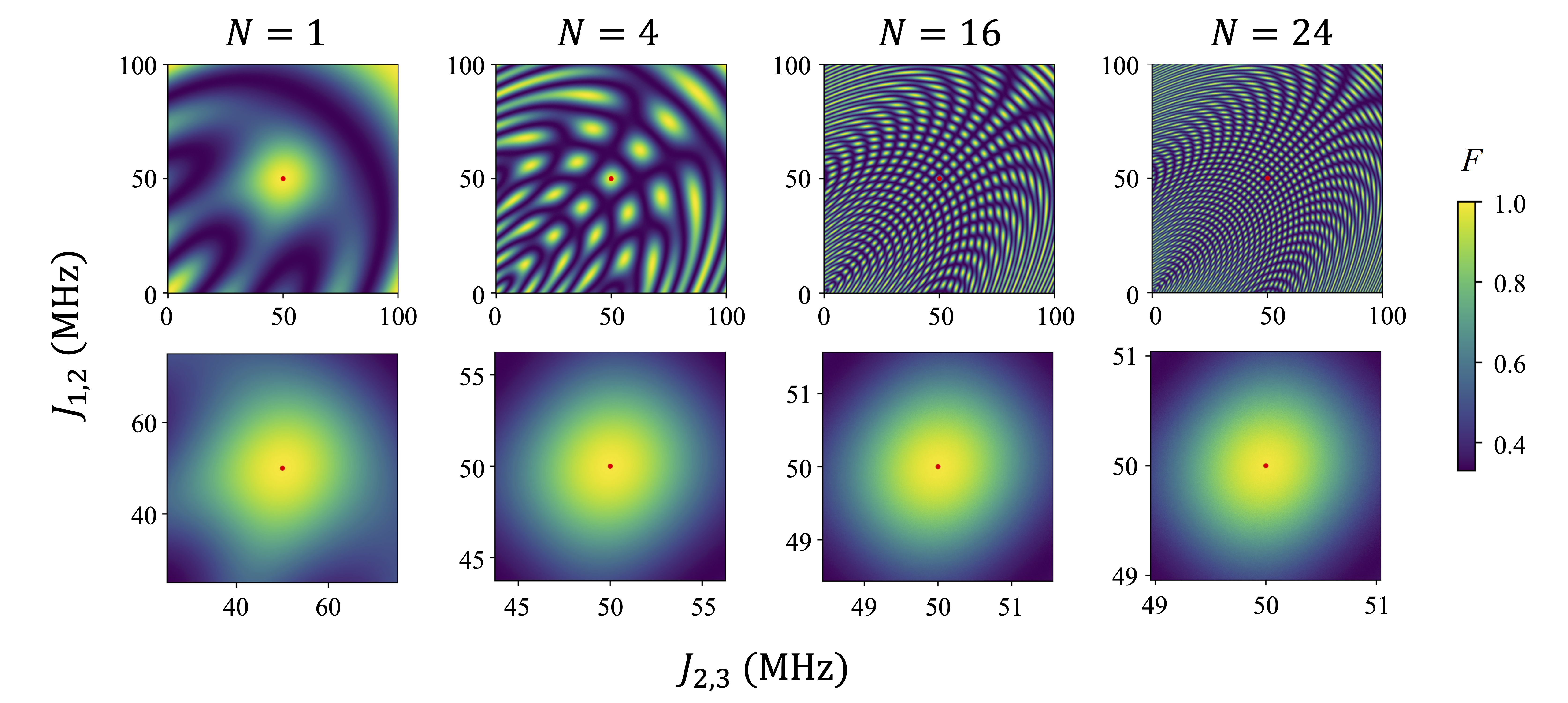}
\caption{\textbf{Scaling of the central calibration peak with} $\boldsymbol{N}$. We consider the calibration of a 2-$J$ rotation $\hat{R}_{\varphi^*}(\theta^*)=\hat{R}_{-\pi/2}(\pi)$, which corresponds to a $\pi$-pulse about the $-\hat{z}$ axis. In this case, the rotation parameters of the pre-calibrated pulse $\hat{R}_{\eta}(\chi)$ are: $\eta=\pi$, and $\chi=\pi$. The plots show the functional dependence of $F$ (Eq.~\ref{eq:F}) on the exchange energies $J_{1,2}$ and $J_{2,3}$ for several different values of $N$. The relationship between $\varphi$ and $\theta$ on $J_{1,2}$ and $J_{2,3}$ are given by Eqs. \ref{eq:Omega}, \ref{eq:cos(phi)}. The bottom row are zoom-ins near the central peaks of the top row. In each plot, the red marker indicates the location of the optimally calibrated 2-$J$ pulse, $\hat{R}_{\varphi^*}(\theta^*)$, which occurs when $J_{1,2}=J_{2,3}=50$ MHz. Evidently, the size of the central peak reduces linearly with increasing $N$.
}
\label{fig:s1}
\end{figure*}

\begin{figure*}[t!]
\renewcommand{\thefigure}{S\arabic{figure}}
\centering
\includegraphics[width=0.99\textwidth]{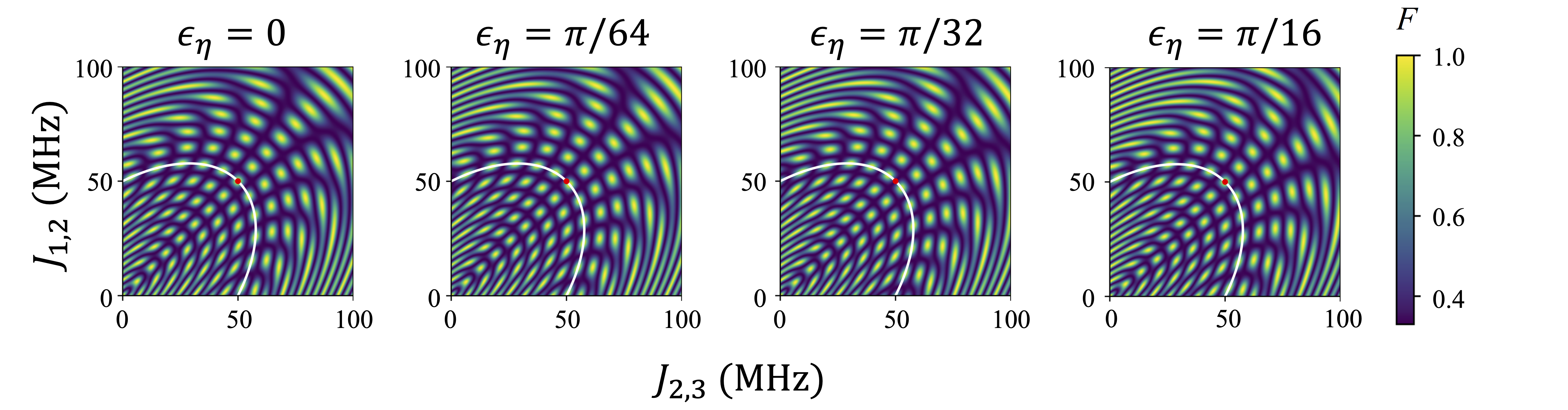}
\caption{\textbf{Effect of errors in} $\boldsymbol{\eta}$. As in Fig.~\ref{fig:s1}, we evaluate $F$ (Eq.~\ref{eq:F}) for a 2-$J$ $\pi$-rotation about the $-\hat{z}$ axis. Here, we fix $N=8$ and consider the effects of errors on the rotation axis of the pre-calibrated pulse $\hat{R}_{\eta}(\chi)$: $\eta=\varphi+\pi/2+\epsilon_{\eta}$. The white curve is a contour of constant $\theta=\theta^*=\pi$. The red marker indicates the location of the optimally calibrated 2-$J$ pulse, $\hat{R}_{\varphi^*}(\theta^*)$, which occurs when $J_{1,2}=J_{2,3}=50$ MHz. Errors, $\epsilon_{\eta}\ne0$, cause the central peak to shift along the contour of constant $\theta$. Along this contour, the interference peaks have a periodicity of $\pi/2N=\pi/16$. Effectively, an error $\epsilon_{\eta}$ causes an equal but opposite error in the calibration of $\varphi$.
}
\label{fig:s2} 
\end{figure*}

\begin{figure*}
\renewcommand{\thefigure}{S\arabic{figure}}
\centering
\includegraphics[width=0.99\textwidth]{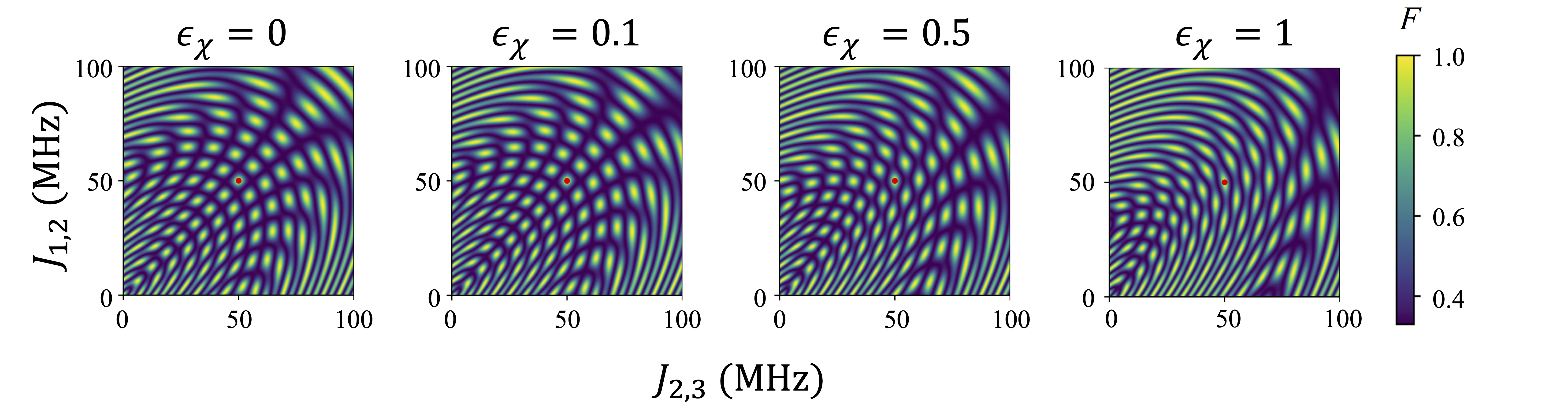}
\caption{\textbf{Effect of errors in} $\boldsymbol{\chi}$. As in Fig.~\ref{fig:s1}, we evaluate $F$ (Eq.~\ref{eq:F}) for a 2-$J$ $\pi$-rotation about the $-\hat{z}$ axis. Here, we fix $N=8$ and consider the effects of errors on the rotation angle of the pre-calibrated pulse $\hat{R}_{\eta}(\chi)$: $\chi=\pi+\epsilon_{\chi}$. The red marker indicates the location of the optimally calibrated 2-$J$ pulse, $\hat{R}_{\varphi^*}(\theta^*)$, which occurs when $J_{1,2}=J_{2,3}=50$ MHz. The central peak is relatively robust to small errors in $\epsilon_{\chi}\ne0$. Moreover, these errors only cause a distortion of the shape of the central peak and not a shift of its mean location. In practice, we leave $\chi$ as a free parameter when fitting to data.
}
\label{fig:s3}
\end{figure*}

When fitting the data, we relate the rotation parameters $\varphi$ and $\theta$ to the exchange energies of the 2-$J$ pulse $\hat{R}_{\varphi}(\theta)$ using using Eqs. \ref{eq:Omega} -- \ref{eq:sin(phi)}. The exchange energies themselves are modeled as having an independent exponential dependence on the exchange gate voltages, described by $J_{i,j}=A\textrm{exp}[B\tilde{V}_{X_{i,j}} + C]$, where $A$, $B$, and $C$ are fit parameters. We find that the validity of this last assumption improves as the range over which the exchange gate voltages are swept decreases. In practice, we only fit the final sweep ($N=24$ for the data presented in the main text) to Eq.~\ref{eq:F}. In the preceeding sweeps, we use a heuristic algorithm to locate the peak: first applying a Gaussian filter to the data, then thresholding at 80\% of the maximum value, and finally computing the centroid, which we identify as the peak. In Figs.~\ref{fig:s1}--\ref{fig:s3}, we use Eq.~\ref{eq:F} to illustrate some important features of the calibration procedure, specifically the scaling with $N$, and the effects of calibration errors on $\eta$ and $\chi$.

We conclude with two remarks on the generality of this procedure. First, while we assumed in this analysis that the pre-calibrated rotation $\hat{R}_{\eta}(\chi)$ was constructed from a single 1- or 2-$J$ exchange pulse, it may be realized as a composite pulse sequence. In that case, an additional error term must be considered in the analysis to account for possible deviations of the rotation axis from the $xz$-plane. However, the advantage of using such composite sequences is that they allow the use of pre-calibrated 1-$J$ gates to tune 2-$J$ rotations about an arbitrary axis in the $xz$-plane. Thus, this procedure directly extends to the more general case. Second, as in robust phase estimation, our calibration procedure is limited to the calibration of rotation angles $\theta$ that are rational multiples of $\pi$ \cite{kimmel2015rpe}. However, it should be possible to extend this procedure to arbitrary $\theta$ by calibrating a sequence of rotations about, say, $\pi/10,\pi/9,\pi/8,\ldots,\pi$, about some axis and then using fits of the data near each of these peaks to perform nonlinear interpolation as is done in the 1-$J$ case \cite{Andrews2019}.

\FloatBarrier

\clearpage

\pagebreak
  
\section{Interleaved blind randomized benchmarking (BRB)}

\begin{figure*}[h]
	\renewcommand{\thefigure}{S\arabic{figure}}
	\centering
	\includegraphics[width=0.99\textwidth]{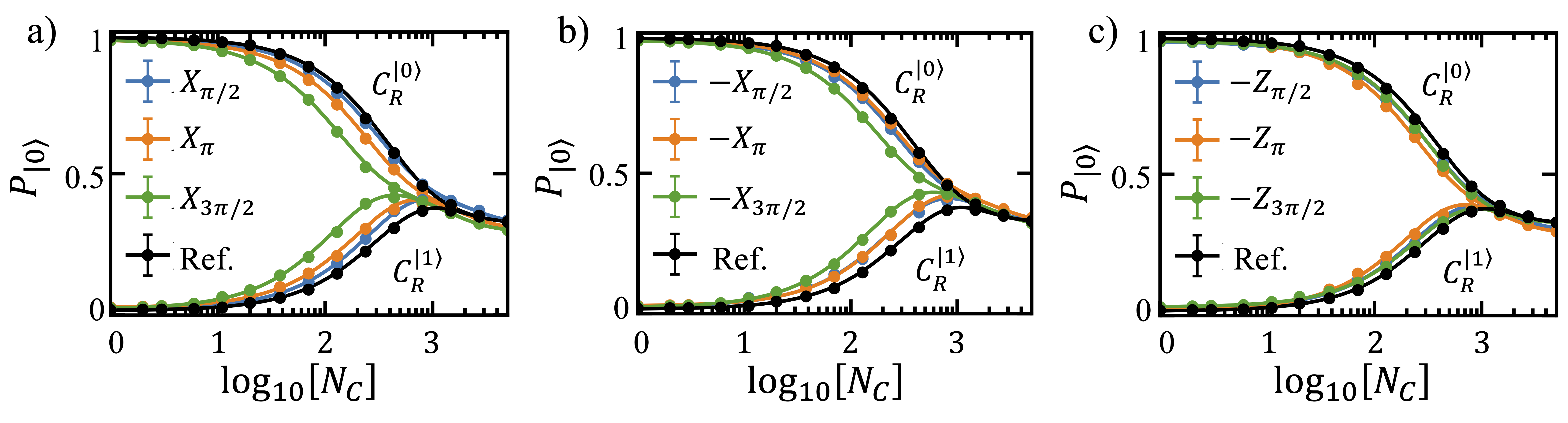}
	\caption{\textbf{Interleaved blind randomized benchmarking}. (a) -- (c) Results of interleaved BRB for 2-$J$ $\pi/2$, $\pi$, and $3\pi/2$ rotations about the $\hat{x}$, $-\hat{x}$, and $-\hat{z}$ axes. The 2-$J$ rotations are interleaved between random sequences of 1-$J$ Clifford gates of depth $N_{C1}$. The reference corresponds to standard BRB using sequences of depth $N_{C1}$ composed of only 1-$J$ Clifford gates. Standard BRB analysis is performed on the resulting data. The interleaved gate errors and interleaved leakage errors are estimated by subtracting the reference sequence errors from the fitted errors of the interleaved sequences. The results of these calculations are summarized in Table~\ref{tbl:irb fidelities}.  Error bars indicate the standard error of the mean probability from 250 sequence repetitions.
	} 
	\label{fig:IBRB}
\end{figure*}

Following the calibration procedure, we tune nine distinct 2-$J$ exchange pulses corresponding to rotations of $\pi/2$, $\pi$, and $3\pi/2$ about each of the axes: $\hat{x}$, $-\hat{x}$, and $-\hat{z}$. The locations of these pulses in bias space are indicated by the colored markers in Fig.~2 of the main text. To evaluate the performance of each pulse, we perform interleaved BRB, which involves interleaving a calibrated 2-$J$ rotation into BRB sequences constructed from 1-$J$ exchange pulses \cite{Andrews2019}. The data are shown in Fig.~\ref{fig:IBRB} and the extracted gate errors are summarized in Table~\ref{tbl:irb fidelities}. Interestingly, the analysis of the interleaved BRB results indicate the average leakage error per 2-$J$ pulse is typically negative, implying that the total leakage error of the sequence is reduced when 2-$J$ pulses are interleaved, despite the increased number of total pulses. This effect is not fully understood, but may arise from a refocusing mechanism similar to that observed when applying dynamical decoupling sequences in triple quantum dot spin qubits \cite{sun2024}.

\begin{table}[t] 
	\begin{tabular}{|c|c|c|c|}
			
			\hline
			Axis                        & Angle    & \begin{tabular}[c]{@{}c@{}}Total Error\\ ($10^{-3}$)\end{tabular} & \begin{tabular}[c]{@{}c@{}}Leakage Error\\ ($10^{-3}$)\end{tabular} \\ \hline
			\multirow{3}{*}{$\hat{x}$}  & $\pi/2$  & 0.240                                                             & -0.095                                                              \\ \cline{2-4} 
			& $\pi$    & 0.794                                                             & -0.018                                                              \\ \cline{2-4} 
			& $3\pi/2$ & 2.48                                                              & 0.044                                                               \\ \hline
			\multirow{3}{*}{$-\hat{x}$} & $\pi/2$  & 0.500                                                             & -0.547                                                              \\ \cline{2-4} 
			& $\pi$    & 0.387                                                             & -0.102                                                              \\ \cline{2-4} 
			& $3\pi/2$ & 1.55                                                              & -0.069                                                              \\ \hline
			\multirow{3}{*}{$-\hat{z}$} & $\pi/2$  & 0.326                                                             & -0.005                                                              \\ \cline{2-4} 
			& $\pi$    & 1.30                                                              & -0.210                                                              \\ \cline{2-4} 
			& $3\pi/2$ & 0.306                                                             & -0.078                                                              \\ \hline
	\end{tabular}
	\caption{\label{tbl:irb fidelities} Summary of 2-$J$ total gate error and leakage errors extracted from interleaved blind randomized benchmarking.}
\end{table}

\FloatBarrier
\bibliography{supplement_library}